\documentclass[10pt,superscriptaddress,aps,pra]{revtex4-1}
\usepackage[utf8x]{inputenc}
\usepackage{amsmath}
\usepackage{amsfonts}
\usepackage{amssymb}

\usepackage{graphicx}

\usepackage[colorlinks=true,linkcolor=black,citecolor=blue,urlcolor=black,pdfpagelabels]{hyperref}

\begin{document} 

\newcommand{\znidaricghost}{}%to make \v{Z}nidaric apper under Z instead of v
\newcommand{\stelmachoghost}{}
\newcommand{\rehacekghost}{}

\title{The local detection method:\\Dynamical detection of quantum discord with local operations}
\author{Manuel Gessner}
\affiliation{QSTAR, INO-CNR and LENS, Largo Enrico Fermi 2, I-50125 Firenze, Italy}
\affiliation{Istituto Nazionale di Ricerca Metrologica (INRiM), I-10135 Torino, Italy}
\author{Heinz-Peter Breuer}
\affiliation{Physikalisches Institut, Albert-Ludwigs-Universit\"at Freiburg, Hermann-Herder-Stra\ss e 3, 79104 Freiburg, Germany}
\author{Andreas Buchleitner}
\affiliation{Physikalisches Institut, Albert-Ludwigs-Universit\"at Freiburg, Hermann-Herder-Stra\ss e 3, 79104 Freiburg, Germany}
\affiliation{Keble College, University of Oxford, Oxford OX1 3PG, UK}
\date{\today}
%\pacs{}

\begin{abstract}
Quantum discord in a bipartite system can be dynamically revealed and quantified through purely local operations on one of the two subsystems. To achieve this, the local detection method harnesses the influence of initial correlations on the reduced dynamics of an interacting bipartite system. This article's aim is to provide an accessible introduction to this method and to review recent theoretical and experimental progress.
\end{abstract}

\maketitle

\tableofcontents

\section{Introduction}
Experimental achievements in the last decades have established the precise quantum control of individual quantum systems \cite{Wineland2013,Haroche2013}. Furthermore, recent efforts are focussed on the assembly and monitoring of interacting quantum systems, with various applications in the context of quantum information \cite{Bloch2008,HartmutReview,Schneider2012,Blatt2012}. The efficient characterization of the underlying quantum states in high-dimensional state spaces, however, remains a challenge due to the large number of parameters. 

One possible strategy for the analysis of systems whose size, complexity or structure is beyond the reach of a detailed microscopic examination is therefore to restrict access to a small, easily controllable subsystem \cite{Haikka2014,GessnerDiss}. By interaction with the remaining system, the locally observable quantities of the subsystem may be able to convey information about the global properties of the interacting system. While in general it is not always clear whether sufficient information about a possibly complex surounding system can be obtained from the few variables of the accessible subsystem, such an approach has proven to be suitable for probing the presence of correlations between the probe and its environment in a variety of situations \cite{Laine2010,GessnerDiss}. In the present article, we review recent progress in the local detection method \cite{Gessner2011,Gessner2013PRA}---an interaction-assisted method, able to reveal quantum discord of the global system through the dynamics of a local subsystem. The method can be implemented when access is restricted to a controllable subsystem, and it has been tested in various different experimental settings. With the help of the examples reviewed in this article we discuss under which physical circumstances a successful local detection based on this method can generally be expected.

The concept of quantum discord can be intuitively understood in terms of local measurements of a bipartite quantum system. Measurements usually induce disturbances of the quantum system under observation \cite{Neumann1955,Lueders1951}. An exception to this textbook rule is found if the system is initially prepared in an eigenstate of the measured observable. More generally, if observable $M$ and quantum state $\rho$ commute, i.e., $[M,\rho]=0$, a non-selective measurement of $M$ will leave the quantum state $\rho$ unchanged \cite{Neumann1955,Lueders1951}. Such a measurement projects the system into the eigenstate $|\varphi_m\rangle$ with probability $p_m=\langle\varphi_m|\rho|\varphi_m\rangle$, where we assume a non-degenerate observable with spectral decomposition $M=\sum_m\lambda_m|\varphi_m\rangle\langle\varphi_m|$. The state at the outcome of the projective measurement is consequently given as
\begin{align}
\Phi(\rho) &= \sum_m p_m|\varphi_m\rangle\langle\varphi_m|\notag\\
&=\sum_m |\varphi_m\rangle\langle\varphi_m|\rho|\varphi_m\rangle\langle\varphi_m|.
\end{align}
In fact, we find that $\Phi(\rho)=\rho$ if and only if $[M,\rho]=0$. Hence, for any given quantum state $\rho$, we can construct a family of observables $M$ which can be measured non-selectively without disturbance. This family is comprised of all observables with the same eigenvectors as $\rho$, assuming no degeneracies.

Let us now consider the case of a bipartite quantum system, described by a tensor product of Hilbert spaces $\mathcal{H}=\mathcal{H}_A\otimes\mathcal{H}_B$. Under which circumstances is it possible to construct \textit{local} observables whose non-selective measurement does not disturb the \textit{total} quantum state $\rho$? The post-measurement state of a non-selective measurement of a \textit{local} observable $M_A\otimes\mathbb{I}_B=\sum_m\lambda_m|\varphi_m\rangle\langle\varphi_m|\otimes\mathbb{I}_B$ is given by
\begin{align}\label{eq.dephasing}
(\Phi\otimes\mathbb{I})\rho&=\sum_m (|\varphi_m\rangle\langle\varphi_m|\otimes\mathbb{I}_B)\rho(|\varphi_m\rangle\langle\varphi_m|\otimes\mathbb{I}_B),
\end{align}
with $\mathbb{I}_B$ the identity on $\mathcal{H}_B$. Again considering only non-degenerate observables, one finds that $(\Phi\otimes\mathbb{I})\rho=\rho$ is indeed equivalent to $[M_A\otimes\mathbb{I}_B,\rho]=0$. The above question can thus be reformulated as: Which quantum states $\rho$ commute with at least one local, non-degenerate observable? Obviously, if systems $A$ and $B$ are completely uncorrelated, i.e., if the total quantum state factorizes as $\rho=\rho_A\otimes\rho_B$, then we can conclude that a family of local observables, e.g. in system $A$, can always be constructed from the eigenvectors of $\rho_A$. The presence of correlations between the two systems, however, changes the situation. 

Only a certain set of quantum states admit the existence of a non-degenerate observable $M_A$, such that $[M_A\otimes\mathbb{I}_B,\rho]=0$. This family is known as the states of zero discord. They can always be written as \cite{Ollivier2001,Henderson2001,Modi2012}
\begin{align}\label{eq.zerod}
\rho_{zd}=\sum_mp_m|\varphi_m\rangle\langle\varphi_m|\otimes\rho_B^m,
\end{align}
where $p_m$ is a probability distribution and $\rho_B^m$ are density operators on system $B$. It is important to note that the $|\varphi_m\rangle$ form an orthonormal basis of $\mathcal{H}_A$ since they are the eigenvectors of the Hermitian operator $M_A$. This distinguishes states of zero discord from separable states with the general form \cite{Werner1989}
\begin{align}
\rho_{sep}=\sum_ip_i\rho_A^i\otimes\rho_B^i,
\end{align}
where the states $\rho_A^i$ are arbitrary and need neither be pure nor orthogonal. Furthermore, unlike entanglement, discord is an asymmetric property, requiring specification of the system which is measured. Throughout this article this will always be system $A$.

Quantum discord therefore characterizes the presence or absence of a local observable which commutes with the full quantum state. Nonzero discord can only be observed in correlated (i.e. not factorizing) quantum states, however, even some separable states exhibit discord. For pure states, the concepts of discord and entanglement coincide. Hence, in general, discord is a concept closely connected to correlations but does not itself measure correlations. In particular, a local operation on system $A$ may change the orthogonality properties of the $|\varphi_m\rangle$ in Eq.~(\ref{eq.zerod}) \cite{Dakic2010,PhysRevA.85.010102,PhysRevLett.107.170502} and thereby create discord without creating correlations \cite{Gessner2012,Lanyon2013}.

The inability to commute with any local observable renders quantum states of nonzero discord furthermore suitable for certain technological tasks \cite{Modi2012}. For instance, a phase shift $\varphi$, imprinted by a local unitary transformation $e^{-iM_A\varphi}\otimes\mathbb{I}_B$ can be estimated with high precision \cite{Helstrom1976} only if the initial quantum state $\rho$ is strongly affected by this transformation \cite{PhysRevLett.112.210401}. Conversely, if $M_A\otimes\mathbb{I}_B$ happens to commute with $\rho$, the state is completely invariant under this transformation and, consequently, an estimation of the phase shift $\varphi$ is impossible. While states of zero discord are insensitive to the action of certain local operators, this can be excluded for all states of nonzero discord, since no local operator commutes with these states \cite{PhysRevLett.112.210401}.

Among other applications \cite{Modi2012,Girolami2013}, discord was further shown to be useful for the distribution \cite{Cubitt2003,PhysRevLett.108.250501,PhysRevLett.109.070501,PhysRevLett.111.230504,PhysRevLett.111.230505,PhysRevLett.111.230506} and activation \cite{PhysRevLett.106.160401,PhysRevLett.106.220403,PhysRevLett.112.140501} of entanglement. To exploit these phenomena experimentally, one needs to first find convenient ways to generate sufficiently robust discordant quantum states \cite{PhysRevA.85.010102,PhysRevLett.107.170502,Lanyon2013,Edo2015,PhysRevLett.115.160503,Carnio2016}. Second, methods to recognize the presence of discord, and perhaps to even quantify discord in experimentally relevant situations are required \cite{PhysRevA.84.032122,Girolami2011,PhysRevA.87.012119,Modi2012}. The local detection method \cite{Gessner2011,Gessner2013PRA}, to be introduced in the next section, is a dynamical method which allows to detect and quantify discord in a bipartite system with limited experimental requirements.

\section{The local detection method}
The efficient detection of properties such as entanglement or discord is a challenging task \cite{Mintert2005,Guhne2009,Horodecki2009}, which usually requires measurements of correlated observables. A popular approach for low-dimensional systems is to first obtain full information about the quantum state, and then to calculate a suitable quantifier based on the measured entries of the density matrix \cite{Haeffner2005,Lanyon2013}. Not only require the results of tomographic reconstructions of quantum states careful statistical analysis \cite{PhysRevLett.114.080403}, their experimental realization soon becomes prohibitively expensive when high-dimensional or multipartite systems are of interest. In these cases, a complete characterization of the full quantum state can no longer be of interest. Alternatively, the measurement of carefully designed observables (such as entanglement witnesses) may reveal the presence of entanglement \cite{PhysRevLett.84.2722,Sorensen2001,PhysRevA.67.022320,PhysRevLett.102.100401,Guhne2009} or discord \cite{PhysRevA.84.032122,Girolami2011,PhysRevLett.110.140501,PhysRevA.87.012119,PhysRevA.90.022305,0953-4075-47-2-025503} without knowledge of the full quantum state. Yet, such procedures are often restricted to Hilbert spaces of a certain dimension and structure (e.g. qubit systems or two-mode systems of continuous variables), cf.~\cite{Gessner2016}. Their implementation furthermore often requires a high degree of control over the full quantum system (or even multiple copies thereof \cite{Walborn2006,PhysRevLett.101.260505,Islam2015}) which is difficult to achieve with increasing system size.

We may also encounter situations in which the experimenter may not even have full access to the complete quantum state of systems $A$ and $B$ but instead may be limited to measurements and operations on the subsystem $A$. This limitation may be due to a fundamental inability to access the second system, when, for instance party $B$ is spatially separated from the experimenter at $A$ or describes degrees of freedom that cannot be measured experimentally. It may also be a deliberate choice such as to restrict the dimension of the quantum system which is to be controlled. In either case one may consider the subsystem $B$ an ancilla system or environment to system $A$. Since the reduced dynamics of system $A$ may be strongly influenced by correlations with system $B$, a dynamical witness for discord may be observable, even by restricting to local measurements of system $A$.

\subsection{Witnessing discord with local operations}
To introduce the basic idea of the local detection method \cite{Gessner2011,Gessner2013PRA} let us recall that states of zero discord are characterized by their invariance under non-selective measurements, i.e., a state $\rho$ has zero discord if and only if there exists a complete set of one-dimensional orthogonal projectors $\boldsymbol{\Pi}=\{\Pi_1,\Pi_2,\dots,\}$, such that $\rho=(\Phi_{\boldsymbol{\Pi}}\otimes\mathbb{I})\rho$, where 
\begin{align}\label{eq.arbdephasing}
(\Phi_{\boldsymbol{\Pi}}\otimes\mathbb{I})\rho = \sum_i(\Pi_i\otimes\mathbb{I}_B)\rho(\Pi_i\otimes\mathbb{I}_B).
\end{align}
This operation~(\ref{eq.arbdephasing}) is a purely local operation on system $A$ and its implementation does not require any knowledge of system $B$. It furthermore describes complete dephasing in the basis $\boldsymbol{\Pi}$ and is therefore called a \textit{local dephasing operation} \cite{Gessner2011}. 

Let us assume the state $\rho$ has zero discord. A local dephasing operation that leaves the state invariant thus exists, but in which particular local basis $\boldsymbol{\Pi}$? Let us first express the full quantum state $\rho$ in terms of families of completely arbitrary local operator bases as $\rho=\sum_\alpha A_\alpha\otimes B_\alpha$. To answer the above question, we study the reduced density matrix of system $A$, which is obtained by performing the partial trace over $B$, after the dephasing operation. We obtain
\begin{align}\label{eq.redstateafterdeph}
\mathrm{Tr}_B\{(\Phi_{\boldsymbol{\Pi}}\otimes\mathbb{I}_B)\rho\}&=\sum_i\sum_\alpha\Pi_iA_\alpha\Pi_i \mathrm{Tr}\{B_\alpha\}\notag\\
&=\sum_{i}p_i\Pi_i,
\end{align}
where $p_i=\sum_\alpha\mathrm{Tr}\{\Pi_iA_\alpha\}\mathrm{Tr}\{B_\alpha\}=\mathrm{Tr}\{(\Pi_i\otimes\mathbb{I}_B)\rho\}$. The invariance property $\rho_A=\mathrm{Tr}_B\{\rho\}=\mathrm{Tr}_B\{(\Phi_{\boldsymbol{\Pi}}\otimes\mathbb{I})\rho\}=\sum_{i}p_i\Pi_i$ shows that the basis $\boldsymbol{\Pi}$ under which a local dephasing operation has no effect on the total quantum state must coincide with the eigenbasis of the reduced density matrix $\rho_A$ \cite{Gessner2011}. For now, we do not consider the case of degeneracies, which may complicate the situation \cite{Gessner2013PRA}.

Hence, one may test for the presence of discord by realizing a local dephasing operation in the eigenbasis of $\rho_A$. The full quantum state is invariant under this operation if and only if it contains no discord. However, a possible change of the full quantum state under the local dephasing operation cannot be directly observed in system $A$: Even if the state contains discord, i.e., it changes under the local dephasing, the resulting reduced density matrix $\rho'_A=\mathrm{Tr}_B\rho'$ with $\rho'=(\Phi_{\boldsymbol{\Pi}}\otimes\mathbb{I}_B)\rho$ will always coincide with $\rho_A$ \cite{Gessner2011}. This can be seen by realizing that the result of Eq.~(\ref{eq.redstateafterdeph}) was derived without making any assumption about the full quantum state $\rho$. We will therefore always observe that $\rho_A=\rho'_A$, regardless of whether $\rho=\rho'$ holds or not.

Moreover, since the local dephasing operation does not act on system $B$, one further finds that also $\rho'_B=\rho_B$ must always hold \cite{Gessner2013PRA}. In fact, we can conclude that if any difference between the original state $\rho$ and the locally dephased reference state $\rho'$ exists, it must be contained in the correlations between the two subsystems---their respective local descriptions are always unchanged by the local dephasing. Does this mean that it is impossible to observe $\rho\neq\rho'$ (which would constitute a witness for discord) by purely local measurements of any of the two systems? A solution can be found by considering the dynamics of this bipartite system. In fact, a change of the correlation properties of the initial state can have a strong observable impact on the reduced dynamics of one of the subsystems. This is especially well known in the theory of open quantum systems \cite{BreuerPetruccione2006}, where the influence of initial system-environment correlations poses a considerable theoretical challenge \cite{Pechukas1994,Alicki1995,Pechukas1995}. Here, however, it can be exploited to reveal a change of the correlations between the subsystems to the local dynamics. Thus, even if $\rho$ and $\rho'$ are indistinguishable to measurements of system $A$ at some initial time $t=0$, they may become locally distinguishable after the two systems have been interacting for a time $t>0$.

\begin{figure}[tb]
\centering
\includegraphics[width=.6\textwidth]{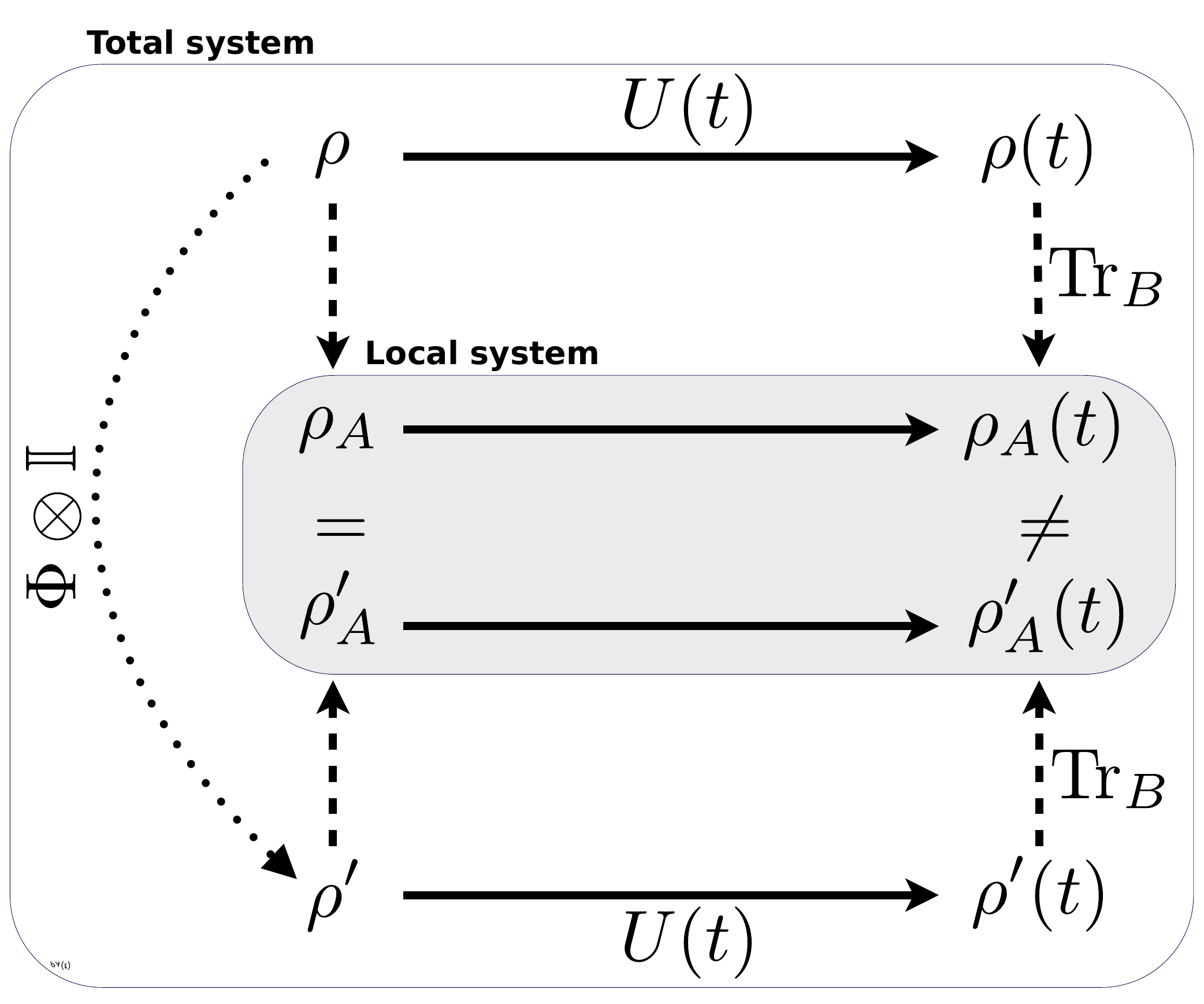}
\caption[Schematic representation of the local detection protocol.]{Schematic representation of the local detection method. A local dephasing operation $\Phi\otimes\mathbb{I}$ produces a reference state $\rho'$, which differs from the original state $\rho$ only in lacking discord. The reduced dynamics of the accessible system $A$ can be strongly influenced by the removal of discord. Hence, at a later time $t$, the system may evolve differently when the initial state $\rho$ is replaced by $\rho'$. If $\rho_A(t)\neq\rho'_A(t)$ is indeed observed, it represents a witness for discord of $\rho$ and the distance among the local states further serves as a quantitative measure of discord. Adapted from \protect\cite{Gessner2014NP}.}
\label{fig.localdetectionprotocol}
\end{figure}

Let us assume that systems $A$ and $B$ are subject to some interaction. For simplicity, we consider a unitary evolution of the composite system \footnote{This assumption is not essential for the local detection method \cite{Gessner2013PRA}.}, such that the evolution in subsystem $A$, given the initial state $\rho$, is governed by
\begin{align}
\rho_A(t)=\mathrm{Tr}_B\{U(t)\rho U^{\dagger}(t)\}.
\end{align}
If the state $\rho$ was subject to local dephasing before the time evolution, the state of system $A$ at time $t$ instead reads
\begin{align}
\rho'_A(t)=\mathrm{Tr}_B\{U(t)\rho' U^{\dagger}(t)\}.
\end{align}
We had already noted that $\rho_A(0)=\rho'_A(0)$, regardless of the properties of $\rho$. However, if we observe
\begin{align}
\rho_A(t)\neq \rho'_A(t)
\end{align}
at a later time $t>0$ we can safely conclude that $\rho\neq\rho'$ which implies the presence of discord in the state $\rho$ \cite{Gessner2011,Gessner2013PRA}. All of the necessary steps, i.e.,
\begin{itemize}
\item Finding the eigenbasis of $\rho_A$
\item Observing the time evolution of $\rho_A(t)$
\item Local dephasing of the initial state in the eigenbasis of $\rho_A$
\item Observing the time evolution of $\rho'_A(t)$
\end{itemize}
can be carried out with strictly local access only to system $A$, whereas no control or even knowledge of system $B$ is required. The local detection protocol is illustrated in the diagram in Fig.~\ref{fig.localdetectionprotocol}.

\subsection{Quantifying discord with local operations}\label{sec.localwitnesstheory}
Detecting the mere presence of discord by following the protocol outlined above does not provide any quantitative information about the discord of the state $\rho$. Considering that discord is a rather ubiquitous phenomenon \cite{Ferraro2010}, it is also relevant to estimate how strongly discordant a given initial state is. Certain quantifiers furthermore allow for an operational interpretation and therefore directly quantify how well a certain quantum information task can be carried out \cite{PhysRevLett.108.250501,PhysRevLett.109.070501,PhysRevLett.106.160401,PhysRevLett.106.220403,PhysRevLett.112.210401,Girolami2013}.

A straight-forward way to quantify discord emerges from the local dephasing operation. From the discussion above, we know that $\rho'$ differs from $\rho$ if and only if $\rho$ contains discord. A simple quantifier of discord is thus given by the \textit{dephasing disturbance} \cite{Luo2008,GessnerDiss}, expressed by the trace distance 
\begin{align}\label{eq.dephasingdisturbance}
D(\rho)=\|\rho-\rho'\|,
\end{align}
where $\|X\|=1/2\mathrm{Tr}\sqrt{X^{\dagger}X}$ \cite{NielsenChuang}. The trace distance has several appealing properties, most notably for our purposes is its contractivity under trace-preserving and positive operations \cite{Ruskai1994}. The unitary time evolution and the partial trace operation are both positive operations (they map positive operators, such as the density operator, to positive operators) \cite{NielsenChuang}. Thus, using the contractivity property we find that the locally observable distance between the reduced density matrices $\rho_A(t)$ and $\rho'_A(t)$ provides a lower bound for the dephasing disturbance \cite{Gessner2013PRA}:
\begin{align}\label{eq.tracedistcontract}
d(t)=\|\rho_A(t)-\rho'_A(t)\|&=\|\mathrm{Tr}_B\{U(t)(\rho-\rho')U^{\dagger}(t)\}\|\notag\\&\leq\|U(t)(\rho-\rho')U^{\dagger}(t)\|\notag\\&=\|\rho-\rho'\|.
\end{align}
The above inequality holds for all $t\geq 0$, hence one may optimize the locally accessible lower bound by observing the time evolution of the local system for as long as possible and then taking the maximum distance \cite{Gessner2014NP}
\begin{align}\label{eq.localmaxdist}
d_{\max}=\max_{t}\|\rho_S(t)-\rho'_S(t)\|\leq \|\rho-\rho'\|.
\end{align}

The above definition assumes that the local dephasing operation is unique, which requires that the eigenbasis of $\rho_A$ is unambiguous. This, however, is not the case if degeneracies are present in the spectrum of $\rho_A$. In these cases the dephasing disturbance does not produce a suitable quantifier of discord \cite{Girolami2011,Campbell2011}. This problem can be circumvented by including a minimization over all possible dephasing bases. Recalling the $\boldsymbol{\Pi}$-dependent definition~(\ref{eq.arbdephasing}) of a general dephasing operation, we introduce the \textit{minimal dephasing disturbance} \cite{Gessner2014EPL,GessnerDiss}
\begin{align}\label{eq.mindephdist}
D_{\min}(\rho)=\min_{\boldsymbol{\Pi}}\|\rho-(\Phi_{\boldsymbol{\Pi}}\otimes\mathbb{I})\rho\|.
\end{align}
This measure in fact quantifies the minimal amount of entanglement which can be activated from discord in a measurement (\textit{minimal entanglement potential}) \cite{PhysRevLett.106.160401,PhysRevLett.106.220403,PhysRevA.88.012117,PhysRevLett.112.140501} when the accessible subsystem is a qubit, i.e.,$\mathcal{H}_A=\mathbb{C}^2$. 

A locally accessible bound for the \textit{minimal dephasing disturbance} can be obtained by dephasing over different local bases instead of just the eigenbasis of $\rho_A$ \cite{Gessner2014EPL}. We introduce
\begin{align}
\rho^{\boldsymbol{\Pi}}_A(t)=\mathrm{Tr}_B\{U(t)(\Phi_{\boldsymbol{\Pi}}\otimes\mathbb{I})\rho U^{\dagger}(t)\}
\end{align}
and the corresponding local trace distance
\begin{align}
d_{\boldsymbol{\Pi}}(t)=\|\rho_A(t)-\rho^{\boldsymbol{\Pi}}_A(t)\|.
\end{align}
A more rigorous bound for discord than Eq.~(\ref{eq.localmaxdist}) is then obtained as \cite{Gessner2014EPL}
\begin{align}\label{eq.minlocalwitness}
d_{\min}(\rho)=\max_t\min_{\boldsymbol{\Pi}}d_{\boldsymbol{\Pi}}(t)\leq D_{\min}(\rho).
\end{align}
To measure the above quantity, one first records the time evolution of $\rho^{\boldsymbol{\Pi}}_S(t)$ for different dephasing bases $\boldsymbol{\Pi}$. At each time $t$, the minimum of all $d_{\boldsymbol{\Pi}}(t)$ is obtained, the minimum being taken over all $\boldsymbol{\Pi}$. Then, within the set of minima one finds the maximum value over all times $t$ to obtain the strongest available local witness. Ideally, the optimization over $\boldsymbol{\Pi}$ should be carried out over all possible bases, which is experimentally impossible. In a realistic situation a systematic sampling over a sufficiently closely spaced grid of basis vectors can yield a good estimate with reasonable overhead, see, e.g., \cite{PhysRevLett.112.140501}.

\subsection{Pure states: Locally accessible lower bound for negativity}\label{sec.negativity}
Let us consider the simple case of a pure state with a controllable qubit subspace, $\mathcal{H}_A=\mathbb{C}^2$. As mentioned before, the concept of discord reduces for pure states to entanglement---in the absence of classical mixing, this is the only form of correlation that can be present. In this case, the dephasing disturbance~(\ref{eq.dephasingdisturbance}) can be evaluated analytically and yields \cite{Gessner2014EPL,GessnerDiss}
\begin{align} 
D(\rho)=\mathcal{N}(\rho),
\end{align}
where $\mathcal{N}$ denotes the \textit{negativity} \cite{Negativity},
\begin{align}\label{eq.neg}
\mathcal{N}(\rho)=\frac{\|\rho^{\Gamma}\|-1}{2},
\end{align}
and $\rho^{\Gamma}$ is the partial transpose of $\rho$ .

On the other hand, the minimal dephasing disturbance~(\ref{eq.mindephdist}) coincides with the minimal entanglement potential, which for pure states also reduces to the negativity \cite{PhysRevA.85.040301}. Hence, in the above scenario, the dephasing disturbance~(\ref{eq.dephasingdisturbance}) coincides with the minimal dephasing disturbance~(\ref{eq.mindephdist}), and the minimum is achieved by dephasing in the eigenbasis of $\rho_A$ \cite{Gessner2014EPL}. Moreover, the local distance~(\ref{eq.localmaxdist}) yields a locally accessible lower bound for the negativity.

\subsection{Efficacy of the local detection method}\label{sec.efficacy}
Discord can ultimately be traced back to those two-body coherences that are present in $\rho$ but are no longer found in $\rho'$, i.e., after local dephasing. Those coherences are not detectable in either of the two subsystems. Therefore, the performance of the local detection method depends crucially on the interacting dynamics between the two subsystems. Its role is to map these initially hidden two-body coherences to measurable elements of the reduced density matrix of system $A$ at a later time. 

Certainly some dynamical processes will work better than others in detecting these correlations. For instance, if no interaction between the two systems were present, i.e., $U(t)=U_A(t)\otimes U_B(t)$, this task could never be achieved \cite{Gessner2013PRA}. In the course of this article we will observe a number of different time evolutions and thereby explore the limitations of the local detection method based on these examples. Let us already mention that early estimates for the efficacy of the method have been obtained based on a formulation in terms of the Hilbert-Schmidt distance, which allows for analytical evaluation of measure-theoretic averages over the unitary group \cite{GessnerPRE}. It was shown that \cite{Gessner2011}
\begin{align}\label{eq.haarmeasureaverage}
\int d\mu(U)\|\mathrm{Tr}_B\{U(\rho-\rho')U^{\dagger}\}\|^2_2=\frac{d_A^2d_B-d_B}{d_A^2d_B^2-1}\|\rho-\rho'\|^2_2,
\end{align}
with the Hilbert-Schmidt norm $\|X\|_2^2=\mathrm{Tr}X^{\dagger}X$, $d_A$ and $d_B$ being the dimensions of the Hilbert spaces $\mathcal{H}_A$ and $\mathcal{H}_B$, respectively, and $d\mu$ representing the Haar measure on the unitary group. These group-theoretic methods further allow for analytical evaluation of the variance corresponding to the above average \cite{Gessner2013PRA}, as well as time-dependent averages with respect to more realistic random matrix ensembles \cite{GessnerPRE}. These results show that the locally observable signal, obtained from a generic dynamical system, is directly proportional to the discord of the initial state. Hence a generic unitary evolution is expected to reveal the quantum discord based on the local detection method. For further details on the Haar-measure integration techniques and additional numerical and analytical studies of the local detection method in this context, we refer to Refs.~\cite{GessnerDA,Gessner2011,Gessner2013PRA,GessnerPRE}. Note, however, that the Hilbert-Schmidt distance is not contractive under positive maps. This can lead to unphysical behavior of Hilbert-Schmidt based quantifiers for discord and related quantities \cite{Ozawa2000,Piani2012}. For this reason, it is generally recommended to use the trace distance instead \cite{Paula2013}.

We also observe that the proportionality factor on the right-hand side of Eq.~(\ref{eq.haarmeasureaverage}) shrinks to zero as $d_B$ increases. This might suggest the conclusion that the signal becomes undetectably small when the observable system is coupled to a truly infinite environment. The result~(\ref{eq.haarmeasureaverage}) however makes statements about generic evolutions which are well represented by the average over all unitaries. Generally speaking, systems that lead to such a dynamics are typically strongly chaotic, and deviations from the average result~(\ref{eq.haarmeasureaverage}) are certainly expected. In common situations one might as well encounter highly non-generic evolutions, in particular in quantum optical systems. Later in this article we will discuss a successful experimental detection of system-environment discord by means of the local detection method where the accessible system couples to an infinite-dimensional, Markovian (memoryless) environment \cite{Tang2014}.

\section{Experiments}
The local detection method has been used to reveal discord in experiments with photons \cite{Tang2014,PhysRevA.90.050301} and trapped ions \cite{Gessner2014NP}. In all experimental applications reported so far, the controllable subspace was two-dimensional, whereas correlations were detected with ancilla systems ranging from two-dimensional systems to continuous variables.

\subsection{Trapped-ion experiment}\label{sec.localdetectionwithion}
%\begin{figure}[tb]
%\centering
%\includegraphics[width=.6\textwidth]{levelschemeNF.pdf}
%\caption[Coupling of electronic qubit and harmonic motion for $^{40}$Ca$^+$ ions.]{The electronic qubit transition can be driven coherently via the quadrupole transition at $729\,$nm. Incoherent readout and dephasing operations are carried out via the $397\,$nm laser. When the detuning of the qubit laser corresponds to multiples of the trap frequency $\omega_x$, the qubit is coupled via the laser to the harmonic ion motion. Figure taken from \protect\cite{Gessner2014NP}.}
%\label{fig.ca40localdetection}
%\end{figure}

An experimental realization of the local detection method was first reported in~\cite{Gessner2014NP}. A single trapped ion is used to simulate both a two-dimensional, locally accessible quantum system by means of its electronic degree of freedom, and a bosonic ancilla system, comprised of the same ion's motional degree of freedom. Since the ion is confined in a harmonic trapping potential, the ancilla system is described by a quantum harmonic oscillator \cite{Wineland1998,Leibfried2003}. When the ion is driven by a laser whose detuning from the ion's resonance transition coincides with the frequency of the harmonic motion, an interaction between the two degrees of freedom, described to a good approximation by the anti-Jaynes-Cummings Hamiltonian
\begin{align}\label{eq.antijcm}
H=\frac{\hbar\Omega\eta}{2}(\sigma_+a^{\dagger}+\sigma_-a).
\end{align}
is evoked \cite{Wineland1998,PhysRevA.46.R6797,Leibfried2003,Blockley,HartmutReview,GessnerDiss}. Here, $a$ and $a^\dagger$ are the bosonic creation and annihilation operators of the harmonic oscillator mode, $\sigma_+$, $\sigma_-$ denote the ladder operators for the electronic qubit system, $\Omega$ denotes the Rabi frequency and $\eta$ is the Lamb-Dicke parameter \cite{Leibfried2003}. In the above expression we assumed, for ease of notation, the Lamb-Dicke regime, i.e., $\eta\sqrt{\langle (a+a^\dagger)^2\rangle}\ll 1$; the analysis of the trapped-ion experiment, however, can be extended beyond this regime and beyond ideal experimental conditions \cite{Gessner2014NP}; for a detailed account see \cite{GessnerDiss}. The Hamiltonian~(\ref{eq.antijcm}) induces a coherent coupling between the states $|g,n\rangle$ and $|e,n+1\rangle$, where $|g\rangle$ and $|e\rangle$ describe the electronic ground- and excited states, respectively, and $|n\rangle$ is a Fock state of the harmonic motion. An ion initially prepared in the state $|g,n\rangle$ hence undergoes a Rabi oscillation of the form
\begin{align}
|\Psi(t)\rangle = U(t)|g,n\rangle=\cos\left(\frac{\Omega_n}{2}t\right)|g,n\rangle+\sin\left(\frac{\Omega_n}{2}t\right)|e,n+1\rangle,
\end{align}
where $U(t)=\exp(-iHt/\hbar)$ and $\Omega_n=\sqrt{n+1}\eta\Omega$.

Initially, the system is prepared by optical pumping of the electronic level to the ground state and laser cooling of the motional degree of freedom, leading to the product state
\begin{align}
\rho_0 = |g\rangle\langle g|\otimes \sum_{n=0}^{\infty}p_n|n\rangle\langle n|.
\end{align}
The thermal populations $p_n=\bar{n}^n/(\bar{n}+1)^{n+1}$ can be given in terms of the mean phonon number $\bar{n}$; see, e.g., \cite{Leibfried2003}. When this initial state is exposed to the laser interaction for a duration $t_0$ (\textit{state preparation}), it evolves as
\begin{align}\label{eq.bluesidebandpulse}
\rho(t_0)&=U(t_0)\rho_0U^{\dagger}(t_0)\notag\\&=\sum_{n=0}^{\infty}p_n\left[\cos\left(\frac{\Omega_n}{2}t_0\right)^2|g,n\rangle\langle g,n|+\sin\left(\frac{\Omega_n}{2}t_0\right)\cos\left(\frac{\Omega_n}{2}t_0\right)|g,n\rangle\langle e,n+1|\right.\notag\\&\hspace{1.5cm}\left.+\sin\left(\frac{\Omega_n}{2}t_0\right)\cos\left(\frac{\Omega_n}{2}t_0\right)|e,n+1\rangle\langle g,n|+\sin\left(\frac{\Omega_n}{2}t_0\right)^2|e,n+1\rangle\langle e,n+1|\right].
\end{align}

The goal is now to detect the presence of discord between the electronic and motional degrees of freedom in the state $\rho(t_0)$. To this end, the local detection method is employed, which allows us to limit experimental access to the electronic degree of freedom. The first task is to obtain the eigenbasis of the reduced density matrix, which determines the basis for the local dephasing operation. By tracing over the motional degrees of freedom (system $B$), we obtain the quantum state of the qubit (system $A$),
\begin{align}\label{eq.reducedstate}
\rho_A(t_0)=\mathrm{Tr}_B\rho(t_0)&=\sum_{n=0}^{\infty}p_{n}\left[\cos\left(\frac{\Omega_n}{2}t_0\right)^2|g\rangle\langle g|+\sin\left(\frac{\Omega_n}{2}t_0\right)^2|e\rangle\langle e|\right],
\end{align}
which is diagonal in the basis $\{|g\rangle,|e\rangle\}$ at all times. 

To detect the discord of the state at time $t_0$, the dynamical evolution is interrupted and a local dephasing is performed by projecting onto the local subspaces, spanned by $|g\rangle$ and $|e\rangle$, respectively. Experimentally this is achieved by inducing a weak ac-Stark shift on the ground state for a well-controlled period of time. To this end, another laser which off-resonantly addresses a transition between the ground state and another short-lived excited state is used. This adds a relative phase shift to any superposition that involves the states $|e\rangle$ and $|g\rangle$. By performing an average over a suitably chosen family of phase shifts, the relative phase relation between the states $|e\rangle$ and $|g\rangle$ can be completely removed, thereby effectively realizing the local dephasing operation \cite{Gessner2014NP,GessnerDiss}. This technique can be combined with local, coherent laser manipulations to achieve dephasing in an arbitrary basis \cite{Gessner2014NP,GessnerDiss}.

The total state after local dephasing is then given as
\begin{align}
\rho'(t_0)&=(\Phi\otimes\mathbb{I})\rho(t_0)\notag\\
&=\sum_{i\in\{e,g\}}(|i\rangle\langle i|\otimes\mathbb{I}_B)\rho(t_0)(|i\rangle\langle i|\otimes\mathbb{I}_B)\notag\\
&=\sum_{n=0}^{\infty}p_{n}\left[\cos\left(\frac{\Omega_n}{2}t_0\right)^2|g,n\rangle\langle g,n|+\sin\left(\frac{\Omega_n}{2}t_0\right)^2|e,n+1\rangle\langle e,n+1|\right].
\end{align}
By construction, $\rho(t_0)$ and $\rho'(t_0)$ only differ inasmuch as $\rho(t_0)$ contains discord while $\rho'(t_0)$ does not, thus, comparison with Eq.~(\ref{eq.bluesidebandpulse}) now allows us to precisely identify those terms that produce the discord in $\rho(t_0)$. As anticipated, these are the two-body coherences $|e,n\rangle\langle g,n+1|$ (and its adjoint counterpart). Since these matrix elements are indeed off-diagonal in both of the sub-systems, any local measurement of the qubit or the ions' motion will be unable to detect their presence. One readily confirms that $\rho'(t_0)$ has the same reduced density matrices as $\rho(t_0)$. Before we proceed to study the signature of discord in the subsequent qubit dynamics, we evaluate the dephasing disturbance, which reads \cite{Gessner2014NP},
\begin{align}\label{eq.tracedistdephdist}
D(\rho(t_0))=\sum_{n=0}^{\infty}p_n\left|\sin\left(\frac{\Omega_n}{2}t_0\right)\cos\left(\frac{\Omega_n}{2}t_0\right)\right|.
\end{align}
As expected, this quantifies precisely the magnitude of the above-mentioned two-body coherences.

\begin{figure}[tbp]
\centering
\includegraphics[width=.6\textwidth]{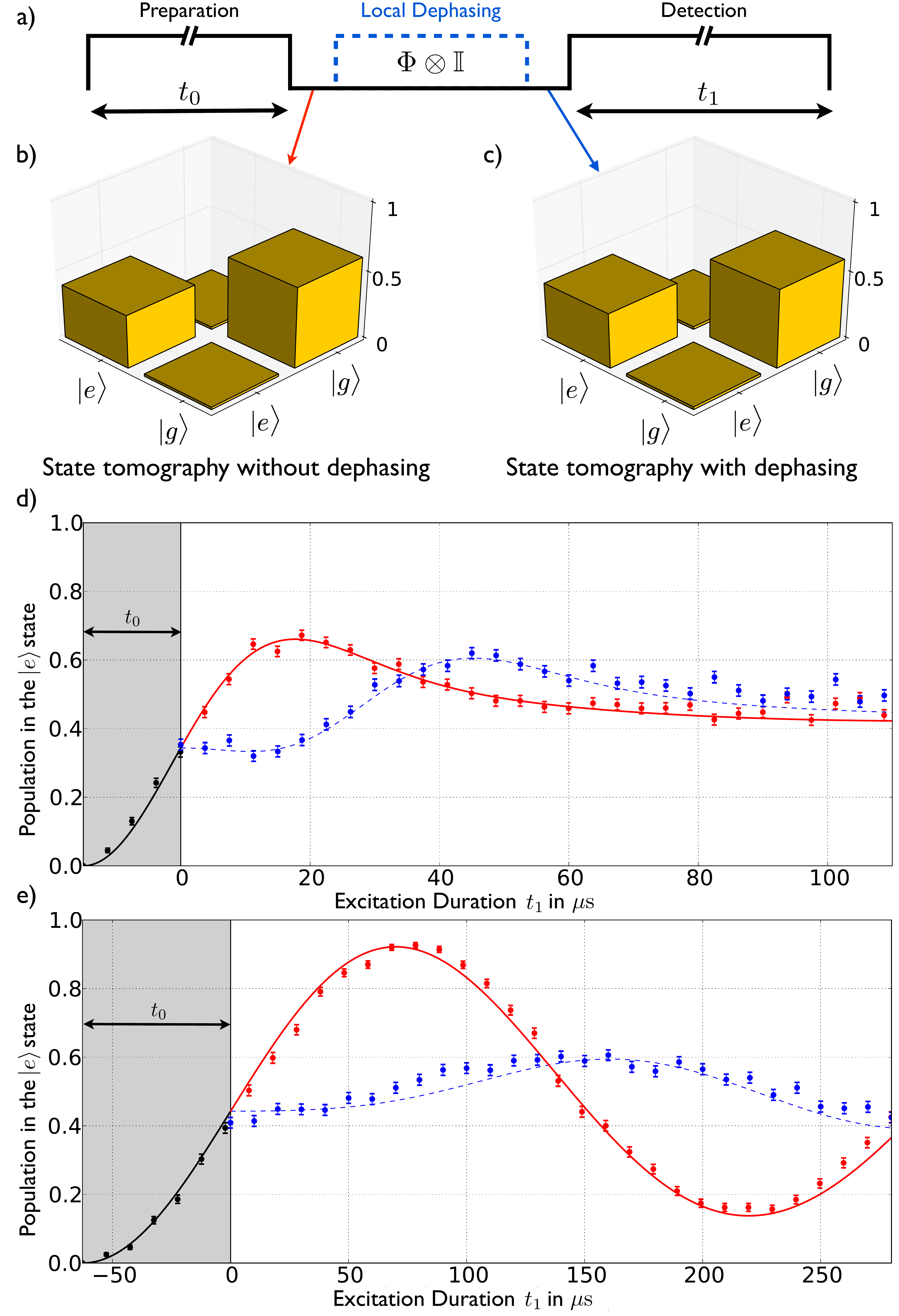}
\caption[Local detection scheme applied to a single trapped ion while driving the first blue sideband.]{The dynamics of the electronic qubit after state preparation is observed with and without dephasing (a). Local state tomography (b) identifies the local basis for the local dephasing operation $\Phi\otimes\mathbb{I}$. Another tomography after dephasing confirms that the local state is initially unchanged (c). The ensuing dynamics (d,e) is, however, strongly influenced by the removal of discord through local dephasing, and any observed difference between the red (unperturbed dynamics) and blue (dynamics after dephasing) data points indicates the presence of discord in $\rho(t_0)$. The average phonon numbers are $\bar{n}=5.9$ in d) and $\bar{n}=0.2$ in e). Figure taken from \protect\cite{Gessner2014NP}.}
\label{fig.ldioncombinedpic}
\end{figure}

Despite being hidden from local measurements, the discord contained in the state $\rho(t_0)$ can be detected at a later time by observing deviating evolutions of the reduced density matrices $\rho_A(t_0+t_1)$ and $\rho'_A(t_0+t_1)$. The state is again subjected to the laser interaction, for a duration $t_1$ (\textit{detection}). The evolution of the unperturbed state was given in Eq.~(\ref{eq.reducedstate}), whereas the dephased state evolves as
\begin{align}
\rho'_A(t_1+t_0)&=\mathrm{Tr}_B\{U(t_1)\rho'(t_0)U^{\dagger}(t_1)\}\notag\\
&=\sum_{n=0}^{\infty}p_n\left[\left(\cos\left(\frac{\Omega_n}{2}t_0\right)^2\cos\left(\frac{\Omega_n}{2}t_1\right)^2+\sin\left(\frac{\Omega_n}{2}t_0\right)^2\sin\left(\frac{\Omega_n}{2}t_1\right)^2\right)|g\rangle\langle g|\right.\notag\\
&\hspace{1.3cm} + \left.\left(\cos\left(\frac{\Omega_n}{2}t_0\right)^2\sin\left(\frac{\Omega_n}{2}t_1\right)^2+\sin\left(\frac{\Omega_n}{2}t_0\right)^2\cos\left(\frac{\Omega_n}{2}t_1\right)^2\right)|e\rangle\langle e|\right].
\end{align}
The difference between the two evolutions can be observed by measuring the excited-state population. In the trapped-ion experiment, this is realized by a highly efficient fluorescence readout method \cite{Wineland1998,Leibfried2003}. We observe the difference
\begin{align}
d_e(t_0,t_1)&=\langle e|\rho_A(t_1+t_0)-\rho'_A(t_1+t_0)|e\rangle\notag\\
&=\frac{1}{2}\sum_{n=0}^{\infty}p_n\sin\left(\Omega_nt_0\right)\sin\left(\Omega_nt_1\right),
\end{align}
where we have used the identity $2\sin\alpha\cos\alpha=\sin 2\alpha$. The fact that both states $\rho_A(t_0)$ and $\rho'_A(t_0)$ are diagonal in the basis $\{|g\rangle,|e\rangle\}$ allows us to determine the trace distance directly from the excited-state deviations as
\begin{align}\label{eq.localtracedist}
d(t_0,t_1)=\|\rho_A(t_1+t_0)-\rho'_A(t_1+t_0)\|=|d_e(t_0,t_1)|.
\end{align}
This quantity is locally measurable in subsystem $A$ and provides a lower bound to the dephasing disturbance~(\ref{eq.tracedistdephdist}), a  global property of the full quantum state. Whenever a nonzero deviation~(\ref{eq.localtracedist}) is observed, we can conclude that $\rho(t_0)$ contained nonzero discord, and the magnitude of the local trace distance further allows for a quantification of the initial discord. 

The experimental protocol and the measured local witness for discord is shown in Fig.~\ref{fig.ldioncombinedpic}. The experiment was performed for two different initial temperatures of the ion's motion. The theoretical description used for the plot includes the effects of experimental imperfections, such as small detunings and fluctuating parameters \cite{Gessner2014NP,GessnerDiss}. For both environmental temperatures, a strong signature of the initial discord is observed.

\begin{figure}[tb]
\centering
\includegraphics[width=.6\textwidth]{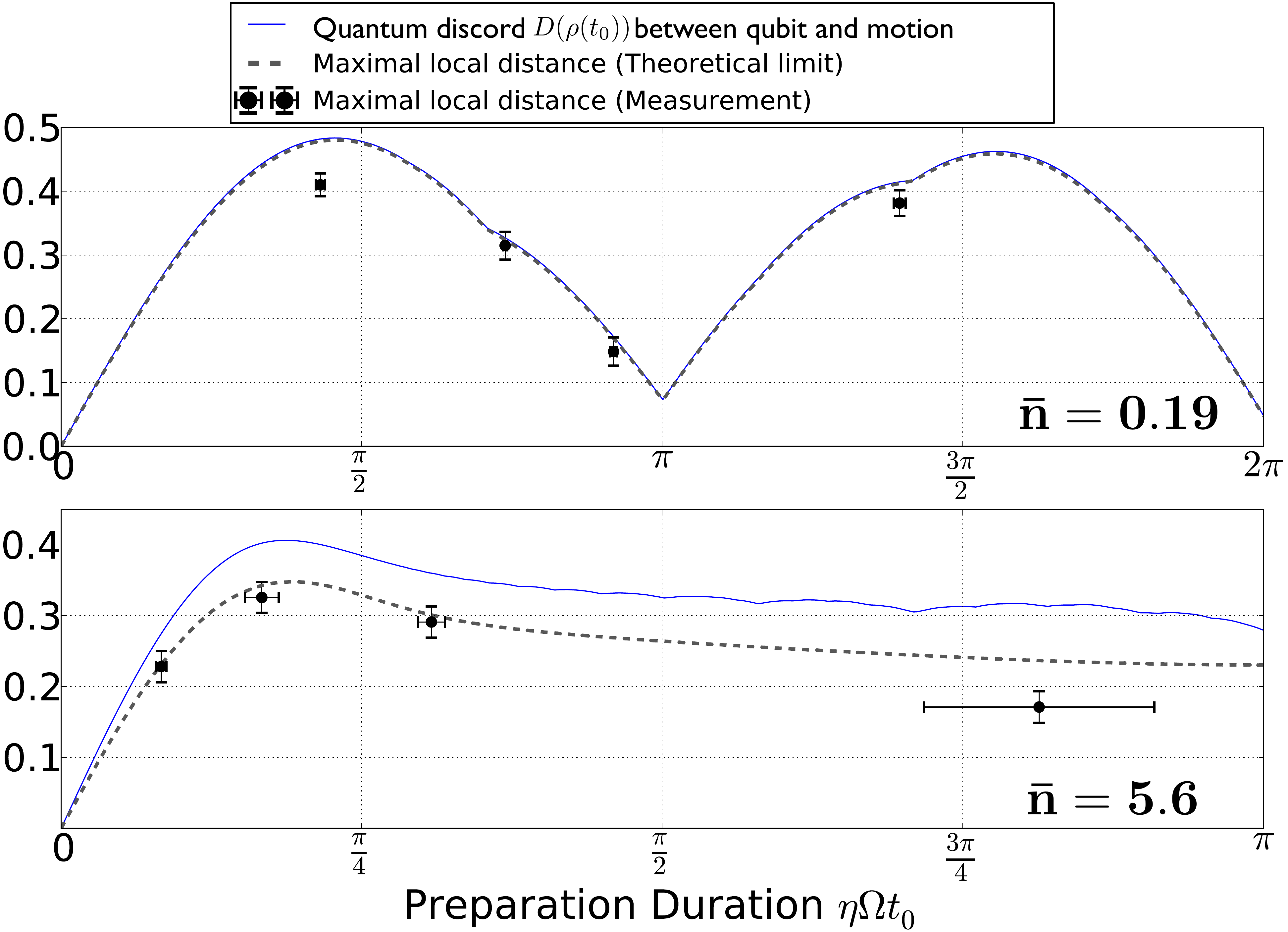}
\caption[Maximum local trace distance provides a lower bound for the dephasing disturbance.]{To obtain the tightest possible bound, the maximum deviation between the local quantum states is taken. In the case of the low-temperature environment the obtained local witness almost saturates the actual distance between the global states. In the higher-temperature case, such a tight estimation is not possible, as shown by the theoretical prediction. Figure taken from \protect\cite{Gessner2014NP,GessnerDiss}.}
\label{fig.maximumlocaldistance}
\end{figure}

Finally, the tightest bound to the dephasing disturbance~(\ref{eq.tracedistdephdist}) is obtained by the largest deviation $d_{\max}$, as introduced in Eq.~(\ref{eq.localmaxdist}). This quantity is plotted for different $t_0$ in Fig.~\ref{fig.maximumlocaldistance}. Comparison with the predictions show that the locally recovered signatures of the initial discord are as large as theoretically possible, and almost saturate the inequality~(\ref{eq.localmaxdist}) in the case of the low-temperature initial state.

The signal of the higher-temperature state is not as pronounced as the one obtained from the low-temperature distribution. The question arises whether the local signal would vanish completely if the temperature was increased even further, as one might expect if the usability of the  local detection method was limited to effectively finite-dimensional environments. This, is however not the case \cite{GessnerDiss}. A simple estimate of the signal for higher temperatures can be obtained by fixing the preparation and detection pulse durations to the value $\Omega_0t_0=\Omega_0t_1=\pi/2$. This leads to the maximum local signal of $1/2$ if the initial state of motion has temperature zero.

\begin{figure}[tb]
\centering
\includegraphics[width=.5\textwidth]{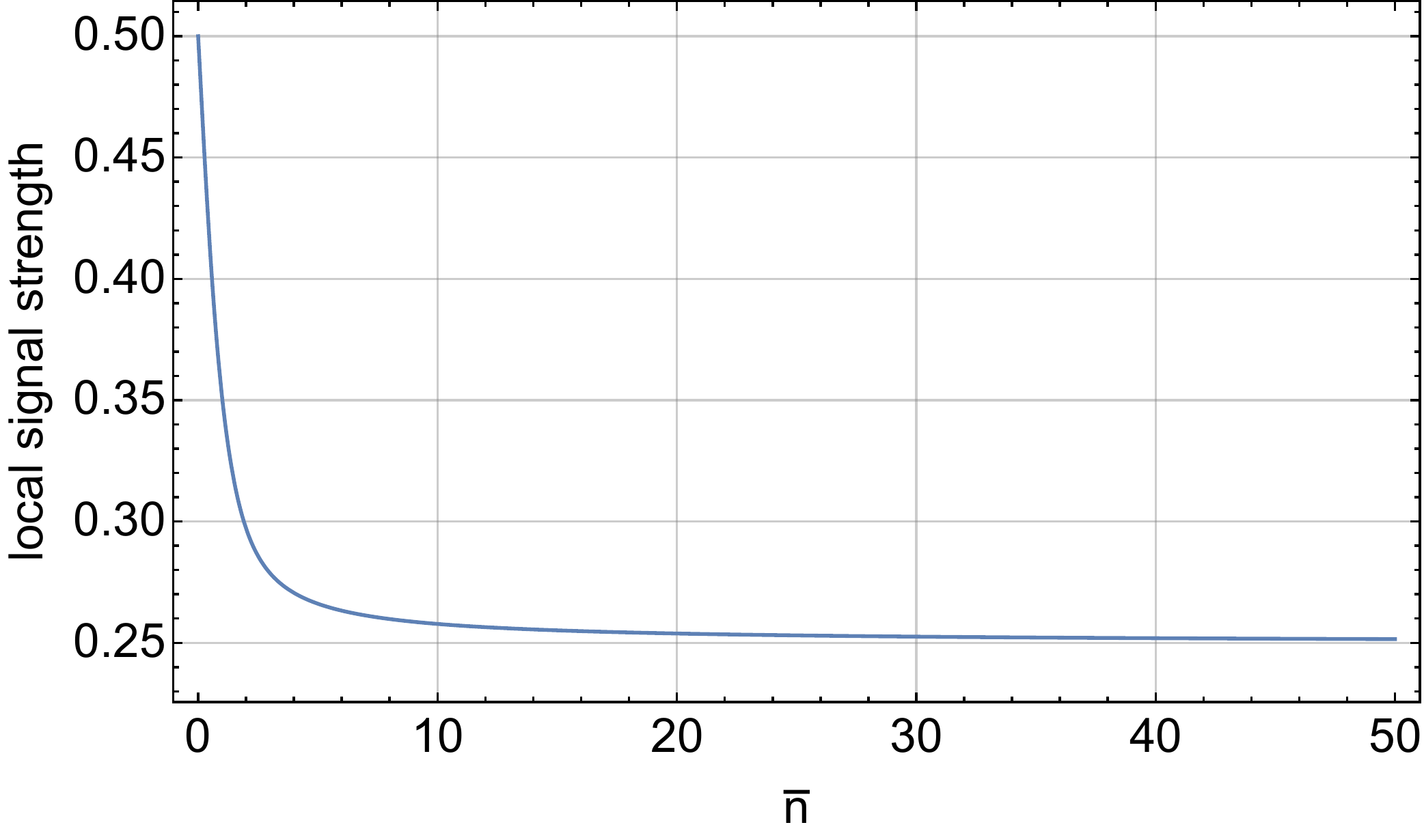}
\caption[Temperature dependence of the local signal.]{Expected local signal of discord as a function of the average thermal phonon number $\bar{n}$, when preparation and detection time are both chosen as $t_0=t_1=\pi/(2\Omega_0)$. The prediction~(\ref{eq.lclsignaltdep}) shows that the local signal remains at a finite value for phonon distributions of much higher temperatures than those realized ($\bar{n}=0.2$ and $\bar{n}=5.9$) in the experiment~\cite{Gessner2014NP}. Figure taken from \cite{GessnerDiss}.}
\label{fig.lclsignaltdep}
\end{figure}

From Eq.~(\ref{eq.localtracedist}), we obtain
\begin{align}\label{eq.lclsignaltdep}
\left. d(t_0,t_1)\right|_{t_0=t_1=\pi/(2\Omega_0)}=\frac{1}{2}\sum_{n=0}^{\infty}\frac{\bar{n}^n}{(\bar{n}+1)^{n+1}}\sin^2\left(\frac{\pi\sqrt{n+1}}{2}\right).
\end{align}
The signal is shown in Fig.~\ref{fig.lclsignaltdep} as a function of $\bar{n}$. After an initial drop, the signal remains close to the finite value of $1/4$ for much higher average phonon numbers than those tested in the experiment. This shows that even for a high-temperature thermal distribution, the local detection method is able to reveal the qubit-motion discord dynamically under an evolution governed by the Hamiltonian~(\ref{eq.antijcm}). We remark that the derivation presented in this section was based on the Lamb-Dicke limit. For sufficiently small values of $\eta$, the expression~(\ref{eq.lclsignaltdep}) still represents a valid approximation for the exact expression even for large values of $\bar{n}$. In particular, values up to $\bar{n}\approx 50$ can be adequately described as long as $\eta\lesssim 0.05$, which applies to the parameter reported in the experiment \cite{Gessner2014NP}. An exact expression for the effective Rabi frequency $\Omega_n$, beyond the Lamb-Dicke limit, can be given in terms of the generalized Laguerre polynomials $L^{(\alpha)}_n(x)$ as $\Omega_n=\eta e^{-\eta^2}(n+1)^{-1/2}L^{(1)}_n(\eta^2)$ \cite{Wineland1998,Leibfried2003,GessnerDiss}. Numerical simulations with the exact expression produce nonzero values of the local signal even when $\bar{n}$ and $\eta$ attain values outside of the Lamb-Dicke limit. For very high values of $\bar{n}$, however, the validity of the effective description of the laser-ion interaction through Eq.~(\ref{eq.antijcm}) reaches its limits, since the fast-moving ion can no longer be laser-addressed with sufficient precision; hence the truly infinite limit $\bar{n}\rightarrow\infty$ cannot be tested with this ansatz.

\subsection{Photonic experiment with continuous-variable ancilla}\label{sec.photonicexperiment}
The first optical realization of the local detection method was reported in \cite{Tang2014}. The accessible system here is represented by a photon's polarization degrees of freedom, which interact with the same photon's frequency degree of freedom when passing through a birefringent material. In contrast to the trapped-ion experiment, the ancilla system is no longer described by a single harmonic oscillator mode, but instead by a continuum of modes.

\begin{figure}[tb]
\centering
\includegraphics[width=.6\textwidth]{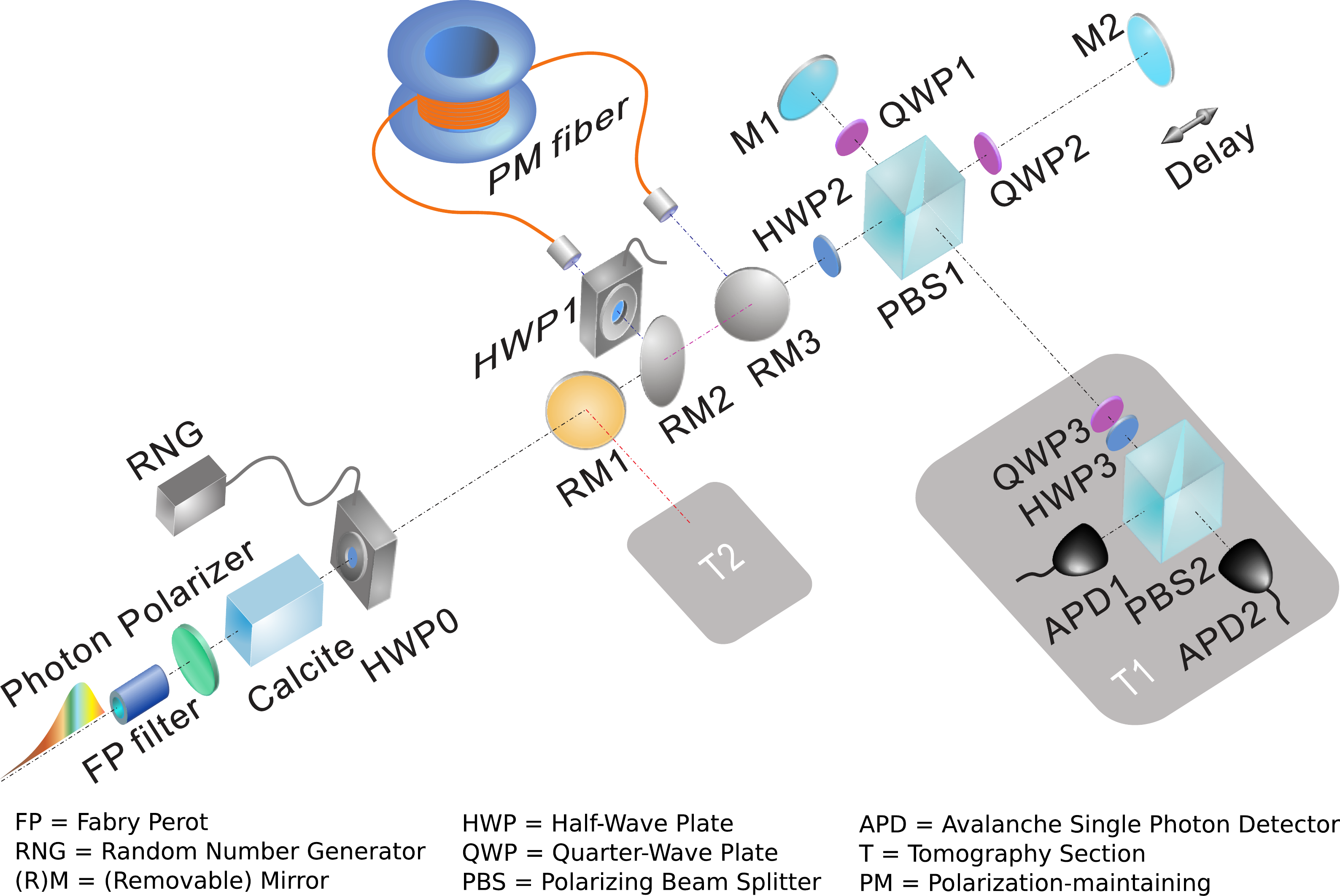}
\caption[Experimental setup for the photonic experiment on local detection of polarization-frequency correlations.]{Experimental setup for local detection with single photons. State preparation is realized by a Fabry-Perot (FP) filter, a polarizer and a calcite crystal, followed by a half-wave plate (HWP0) with a random orientation to scramble the local basis. A removable mirror (RM1) can be inserted to send the photon into the local state tomography unit (T2) which is used to obtain the full quantum state of polarization. When the mirrors RM1-3 are removed, the state is sent into the Michelson interferometer to generate an interacting dynamics between polarization and frequency, by adjusting the position of mirror M2. This evolution is followed by another tomography section T1 to measure the local dynamics. To reveal discord, local dephasing is realized by placing RM2 and RM3 to send the photon through a long polarization-maintaining (PM) fiber which removes all discord, when the local eigenbasis has been mapped onto its principal axes by means of HWP1. This may affect the ensuing polarization dynamics observed in T1 which would constitute a witness for discord. Figure adapted from \protect\cite{Tang2014}.}
\label{fig.photonsetup}
\end{figure}

The experimental setup is summarized in Fig.~\ref{fig.photonsetup}. Initially single photons are created in the quantum state
\begin{align}
\rho_{pi}&=\sum_{\omega}\Delta\omega G(\omega)\left(\frac{1}{2}|H,\omega\rangle\langle H,\omega| +\beta e^{i\varphi}|H,\omega\rangle\langle V,\omega| \right.\notag\\
&\left.\hspace{2.7cm}+\:\beta e^{-i\varphi}|V,\omega\rangle\langle H,\omega|+\frac{1}{2}|V,\omega\rangle\langle V,\omega|\right),
\end{align}
where the mixed frequency distribution is described by a Lorentzian line shape,
\begin{align}\label{eq.lorentzian}
G(\omega)=\frac{1}{\pi}\frac{\delta\omega}{\delta\omega^2+(\omega-\omega_0)^2}.
\end{align}
Here, we have discretized the frequency space by introducing a small frequency interval $\Delta\omega$; later on we will consider the continuum limit $\Delta\omega\rightarrow 0$. A basis for the polarization state is defined by the states $\{|H\rangle,|V\rangle\}$, describing horizontal and vertical polarization, respectively. When passing through a birefringent material, states with a specific polarization direction travel with a modified velocity, and, due to a different dwell time inside the material, experience a modified phase shift. Formally, we find
\begin{align}
U_{\mathrm{c}}(t):\begin{cases}|H,\omega\rangle\rightarrow|H,\omega\rangle\\
|V,\omega\rangle\rightarrow e^{-i\omega t}|V,\omega\rangle\end{cases},
\end{align}
where the dwell time is given as $t=\Delta n_{\rm{c}} L/c$, with the speed of light $c$, the length $L$ of the crystal and, the birefringence $\Delta n_{\mathrm{c}}$ describing the difference between the refractive indices for the two polarization directions. This evolution produces the correlated states
\begin{align}\label{eq.rhosephotoncal}
\rho&=U_{\mathrm{cal}}(t)\rho_{pi}U^{\dagger}_{\mathrm{cal}}(t)\notag\\
&=\sum_{\omega}\Delta\omega G(\omega)\left(\frac{1}{2}|H,\omega\rangle\langle H,\omega|+\beta e^{i(\omega t+\varphi)}|H,\omega\rangle\langle V,\omega|\right.\notag\\
&\hspace{2.7cm}\left.+\:\beta e^{-i(\omega t+\varphi)}|V,\omega\rangle\langle H,\omega|+\frac{1}{2}|V,\omega\rangle\langle V,\omega|\right).
\end{align}
In the experiment, the initial phase $\varphi$ was chosen such that $\varphi=-\omega_0t$.

To reveal the discord of $\rho$ with the local detection method, one first determines the reduced state of the accessible system, which in this case is the qubit state
\begin{align}\label{eq.reducedstatephot}
\rho_A=\begin{pmatrix}1/2 & \beta C(t)\\
\beta C(t)&1/2\end{pmatrix},
\end{align}
with the real-valued function
\begin{align}
C(t)=\sum_{\omega}\Delta\omega G(\omega)e^{i(\omega-\omega_0)t}.
\end{align}
Experimentally, this state is obtained by inserting the removable mirror RM1, and using the tomography section T2, as is pictured in Fig.~\ref{fig.photonsetup}. In contrast to the trapped-ion experiment, where the local eigenbasis was always given by the computational basis, here, the eigenvectors are given by $|\pm\rangle=\frac{1}{\sqrt{2}}(|H\rangle \pm|V\rangle)$. Note that the local eigenbasis is first hidden by a random local unitary basis rotation of the qubit, controlled by the half-wave plate HWP0 in Fig.~\ref{fig.photonsetup}. While this step renders the experimental detection of discord more challenging, it does not affect the theoretical treatment, since it can be accounted for by an effective redefinition of the local basis, which does not alter the correlation properties.

To realize a dephasing of the qubit experimentally, the eigenstates of the photon are mapped onto the principal axes of a long polarization-maintaining fiber by means of HWP1 after removing the mirror RM1, and placing the mirrors RM2 and RM3; see Fig.~\ref{fig.photonsetup}. The small birefringence of the fiber leads to an effective dephasing after a sufficiently long interaction time, such that the locally dephased reference state
\begin{align}\label{eq.rhoseprime}
\rho'&=(\Phi\otimes\mathbb{I})\rho\\
&=\sum_{i\in\{+,-\}}(|i\rangle\langle i|\otimes\mathbb{I}_B)\rho(|i\rangle\langle i|\otimes\mathbb{I}_B)\notag\\
&=\frac{1}{2}\sum_{\omega}\Delta\omega G(\omega)\Big[|H,\omega\rangle\langle H,\omega|+|V,\omega\rangle\langle V,\omega|\notag\\
&\hspace{2.7cm}+\:\beta(e^{i(\omega-\omega_0) t}+e^{-i(\omega-\omega_0) t})|H,\omega\rangle\langle V,\omega|\notag\\
&\hspace{2.7cm}+\:\beta(e^{-i(\omega-\omega_0) t}+e^{i(\omega-\omega_0) t})|V,\omega\rangle\langle H,\omega|\Big],\notag
\end{align}
is created \cite{Tang2014}.

One may again confirm that the reduced density matrices describing polarization and frequency degrees of freedom coincide for $\rho$ and $\rho'$ \cite{GessnerDiss}. The dephasing disturbance~(\ref{eq.dephasingdisturbance}) is evaluated in the continuum limit $\Delta\omega\rightarrow 0$, i.e., $\sum_{\omega}\Delta\omega\rightarrow\int d\omega$, and yields \cite{Tang2014,GessnerDiss}
\begin{align}\label{eq.totaldistance}
D(\rho)=\frac{\beta}{2}\int d\omega G(\omega)\left|e^{i(\omega-\omega_0) t}-e^{-i(\omega-\omega_0) t}\right|.
\end{align}

For the dynamical detection of the discord~(\ref{eq.totaldistance}), the Michelson delay setup, consisting of HWP2 with tunable angle $\eta/2$, and a polarizing beam splitter (PBS1) is used. Theoretically the realized dynamics is equivalent to the one described in the birefringent material, since, again one of the two polarization states acquires a modified phase shift due to a different dwell time. The dwell time $\tau=2x/c$ in the Michelson setup is determined by the delay $x$ of the mirror M2. 

\begin{figure}[tb]
\centering
\includegraphics[width=.6\textwidth]{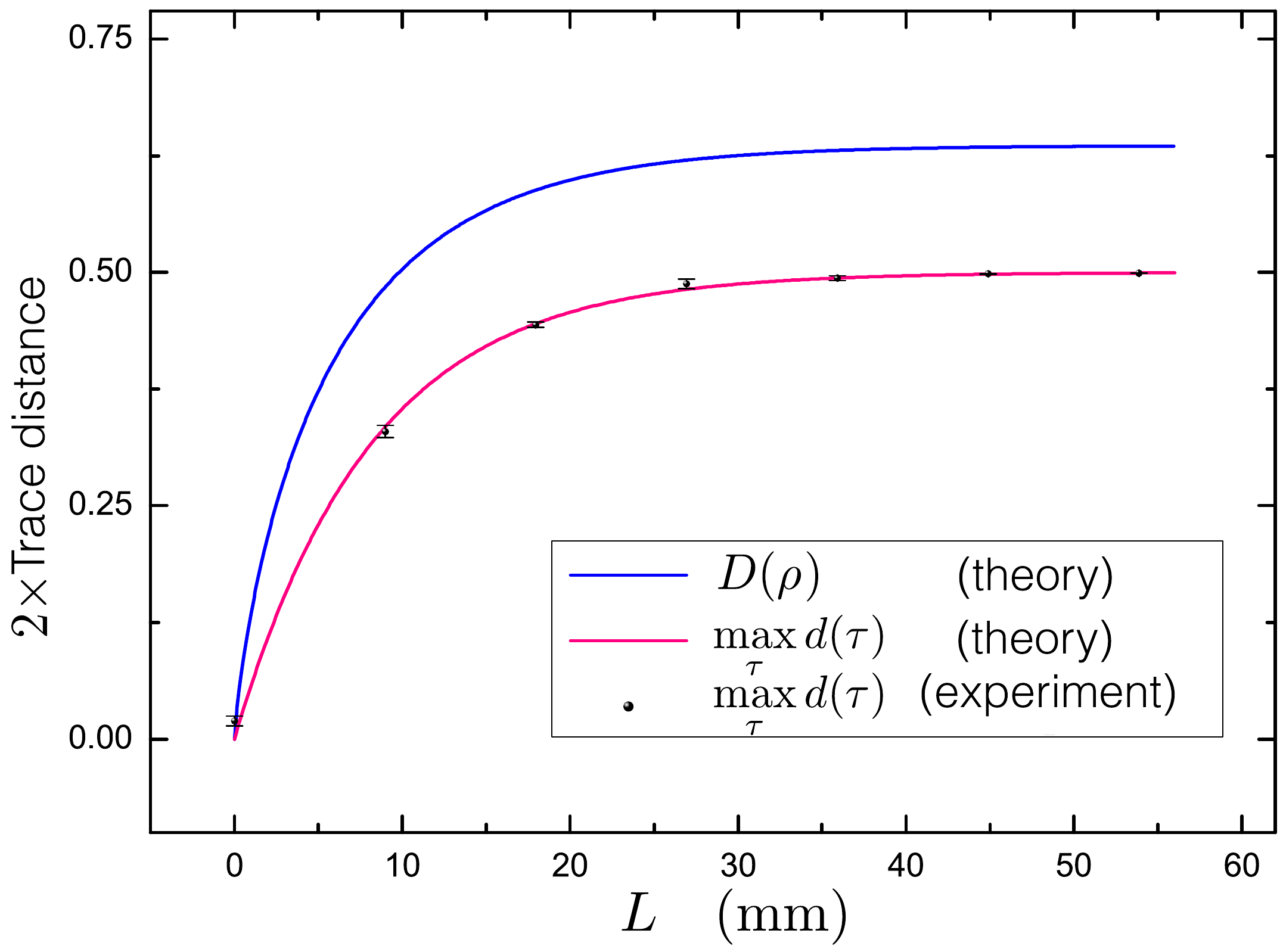}
\caption[Maximum local distance bounds the total distance in the photonic experiment.]{The maximum local trace distance~(\ref{eq.maxlocaldist}), $\mathrm{max}_{\tau}d(\tau)$, provides a lower bound for the total trace distance~(\ref{eq.totaldistance}), $D(\rho)$, for all states~(\ref{eq.rhosephotoncal}). While the full amount of the correlations cannot be revealed locally, the theoretical limit is reached with high precision. Figure adapted from \protect\cite{Tang2014}.}
\label{fig.maxdistphotons}
\end{figure}

The resulting local trace distance can be shown to be independent of $\eta$, and reads in the continuum limit \cite{Tang2014},
\begin{align}\label{eq.localdistance}
d(\tau)&=\frac{\beta}{2} \left|\int d\omega G(\omega)
 \left( e^{i(\omega-\omega_0) t} - e^{-i(\omega-\omega_0) t} \right) e^{i\omega\tau}
 \right|\notag\\
&=\frac{\beta}{2}\left|e^{-\delta\omega|t+\tau|}-e^{-\delta\omega|t-\tau|}\right|.
\end{align}
Its maximum value
\begin{align}\label{eq.maxlocaldist}
\max_{\tau}d(\tau)=\frac{\beta}{2}(1-e^{-2\delta\omega t}),
\end{align}
produces a bound to the dephasing disturbance~(\ref{eq.totaldistance}), as depicted in Fig.~\ref{fig.maxdistphotons} for different values of $t$. The figure shows a strong dynamical signal of the discord, as well as excellent agreement between experiment and theory.

The dynamics of the qubit system, evoked through interaction with its frequency degrees of freedom, could be described as pure dephasing. No excitations are exchanged between the polarization and frequency degrees of freedom. Instead the presence of correlations between these two subsystems effectively leads to the decay of coherences in the qubit system. This dynamics if furthermore completely irreversible, and can thus be considered as Markovian, i.e., memoryless open-system evolution \cite{Liu2011,RevModPhys.88.021002}, with the time scale of the decay being determined by the width $\delta\omega$ of the initial frequency distribution \cite{BreuerPetruccione2006}. Hence, in the case of a purely dephasing coupling between system and environment, the reported experiment demonstrates the applicability of the local detection method to reveal initial system-environment discord even if the environment is completely memoryless \cite{GessnerDiss}.

\subsection{Photonic experiment with discrete-variable ancilla}
In another photonic experiment, discord between the polarization and momentum degrees of freedom of a photon, created in the process of parametric downconversion, was detected using the local detection method \cite{PhysRevA.90.050301}. The momentum space is restricted here to two possible channels, denoted by $|0\rangle$ and $|1\rangle$. Hence, the ancilla system is effectively described by a discrete two-dimensional state space. This distinguishes the setup from the two experiments discussed before, where the ancilla was described by single- \cite{Gessner2014NP} and multi-mode \cite{Tang2014} harmonic oscillators, respectively.

The state of one of the two photons which are created during parametric downconversion is, after suitable manipulations by a double-slit and a tunable half-wave plate, described by
\begin{align}\label{eq.rk2nzd}
\rho=\lambda|H\rangle\langle H|\otimes |0\rangle\langle 0|+(1-\lambda)|\theta\rangle\langle \theta|\otimes|1\rangle\langle 1|,
\end{align}
with $|\theta\rangle=\cos\theta|H\rangle+\sin\theta|V\rangle$. This state contains no discord only if the angle $\theta$ is chosen such that $|\theta\rangle=|H\rangle$ or $|\theta\rangle=|V\rangle$. For all other values the state contains discord since $|\theta\rangle$ and $|H\rangle$ are neither orthogonal nor parallel. As a side remark, we note that this type of discordant state can be (and was in fact) generated via a local operation from a zero-discord state, which is reflected by its low correlation rank \cite{Gessner2012}.

The accessible system is given, again, by the polarization degree of freedom, and the associated qubit state is obtained through full state tomography. The local dephasing operation is then implemented using suitably adjusted polarizers. For the dynamical detection of discord a unitary evolution is realized: By inserting a relative phase shift between the two polarization states in only one of the two momentum channels, an effective interaction between the polarization and the momentum degrees of freedom is mediated. This effectively leads to pure dephasing of the polarization state and is close in spirit to the dynamics described in the previous experiment. Using these ingredients and following the local detection protocol, the discord of the initial states~(\ref{eq.rk2nzd}) was successfully revealed in the experiment. 

When $|\theta\rangle=|V\rangle$, however, the initial state does not contain discord. Nevertheless it is still classically correlated since it cannot be written as a factorizing product state. The deviation from a product state can also be dynamically revealed by observing an increase of the trace distance above its initial value, when comparing the evolution of the unperturbed state with the evolution after an arbitrary local operation \cite{Laine2010,RevModPhys.88.021002}; for earlier experiments see \cite{Li2011,Smirne2011,PhysRevA.88.012108}. In the experiment, this is realized whenever the local detection method did not produce a witness for discord. In this case, a local unitary operation, implemented through suitably placed half-wave plates, produces a reference state, such that the presence of correlations in the initial state was detected by means of an increase of the trace distance above its initial value, except when the initial state is indeed factorized \cite{PhysRevA.90.050301}. This way, the resulting two-step protocol, comprised of a combination of the local detection method \cite{Gessner2011,Gessner2013PRA} and the trace distance witness for initial correlations \cite{Laine2010}, is able, on the one hand, to detect discord, and on the other hand, to reveal classical correlations in the absence of discord \cite{PhysRevA.90.050301}.

\section{Theoretical studies}
Aside from providing an experimentally convenient method for the detection of discord, the local detection method may also be helpful to gain insight into the impact of correlations and discord through theoretical studies of interacting systems. Such studies may further provide a useful characterization of the local detection method itself, by indicating the conditions under which the presence of discord can be successfully revealed through the local dynamics.

\subsection{Dynamical single-spin signature of a quantum phase transition}
\label{sec.epl}
A special situation arises when the state $|\Psi\rangle$ to which the local detection method is applied, is at the same time an eigenstate of the Hamiltonian which governs the interacting time evolution of the system \cite{Gessner2014EPL}. In this case, the dynamics without dephasing is trivially constant and any dynamics that arises after the local dephasing operation constitutes a witness for discord. Additionally, the local trace distance provides a lower bound to the bipartite entanglement of the system (the bipartition is defined by the subspace on which the dephasing was implemented and the rest of the system). Such a situation is particularly interesting when $|\Psi\rangle$ is chosen as the ground state of a many-body system \cite{Gessner2014EPL}, since its quantum correlations can disclose information about the existence of a quantum phase transition \cite{Osterloh2002,PhysRevA.66.032110}.

The local detection method has been applied in a theoretical study to reveal the quantum correlations of the ground state, as well as quantum discord of finite-temperature thermal equilibrium states in the context of a quantum phase transition \cite{Gessner2014EPL}. The properties of the ground state are often regarded as a principal indicator of critical phenomena, since quantum phase transitions are defined as abrupt qualitative changes of the ground state as a function of some external control parameter \cite{Sac99}. By using the local detection method to reveal entanglement properties of the ground state to the dynamics of a single spin, we connect these ground-state properties to the entire excitation spectrum, which is relevant for the dynamical evolution of the spin.

The system studied here is the Ising model with variable interaction range \cite{Dyson1969,PorrasCiracSpins,Tobi,Britton,Jurcevic2014,Richerme2014,PhysRevB.93.155153,GessnerDiss}
\begin{align}\label{eq.spinchainepl}
H_\alpha = -\sum_{\substack{i,j=1\\(i < j)}}^N \frac{J_0}{|i-j|^\alpha} \sigma_x^{(i)} \sigma_x^{(j)} - B \sum_{i=1}^N \sigma_y^{(i)},
\end{align}
where $\sigma_k^{(i)}$ describe the Pauli matrices for spin $i$ with $k\in\{x,y,z\}$, $J_0$ determines the strength of the spin-spin interaction, which stands in competition with the transverse external field of strength $B$. The ground state provides an intuitive understanding of the quantum phase transition: For small values of $B$ the contribution of the internal spin-spin interaction dominates and the relative orientation of the spins is chosen such that the associated potential energy is minimized, thereby describing a ferromagnetic state for $J_0>0$. When $B$ is increased above a critical value, which depends on $\alpha$ and $J_0$, the spins tend to align along the direction of the external fields and the system describes a paramagnet.

\subsubsection{Local bound for the ground-state negativity}
To apply the local detection method to the ground state of the above system, we consider the left-most spin as the controllable subsystem and consider the ensemble of all other spins as the inaccessible ancilla. This reduces the state space from an exponentially large dimension of $2^N$ to the easily manageable size of a qubit. 

The initial state $|\Psi_0\rangle$, being the ground state of $H_\alpha$, does not evolve in time. However, by applying the local dephasing operation to the state $|\Psi_0\rangle$ excited states are incoherently populated, which does no longer necessarily result in a time-invariant state. Any time evolution is, according to the local detection method, a witness for discord, and, since the state is pure, in this case also a witness for entanglement. As described in Section~\ref{sec.negativity}, local measurements of the single spin dynamics can be used to obtain a lower bound for the negativity (a simple entanglement measure) of $|\Psi_0\rangle$.

\begin{figure}[tb]
\centering
\includegraphics[width=.5\textwidth]{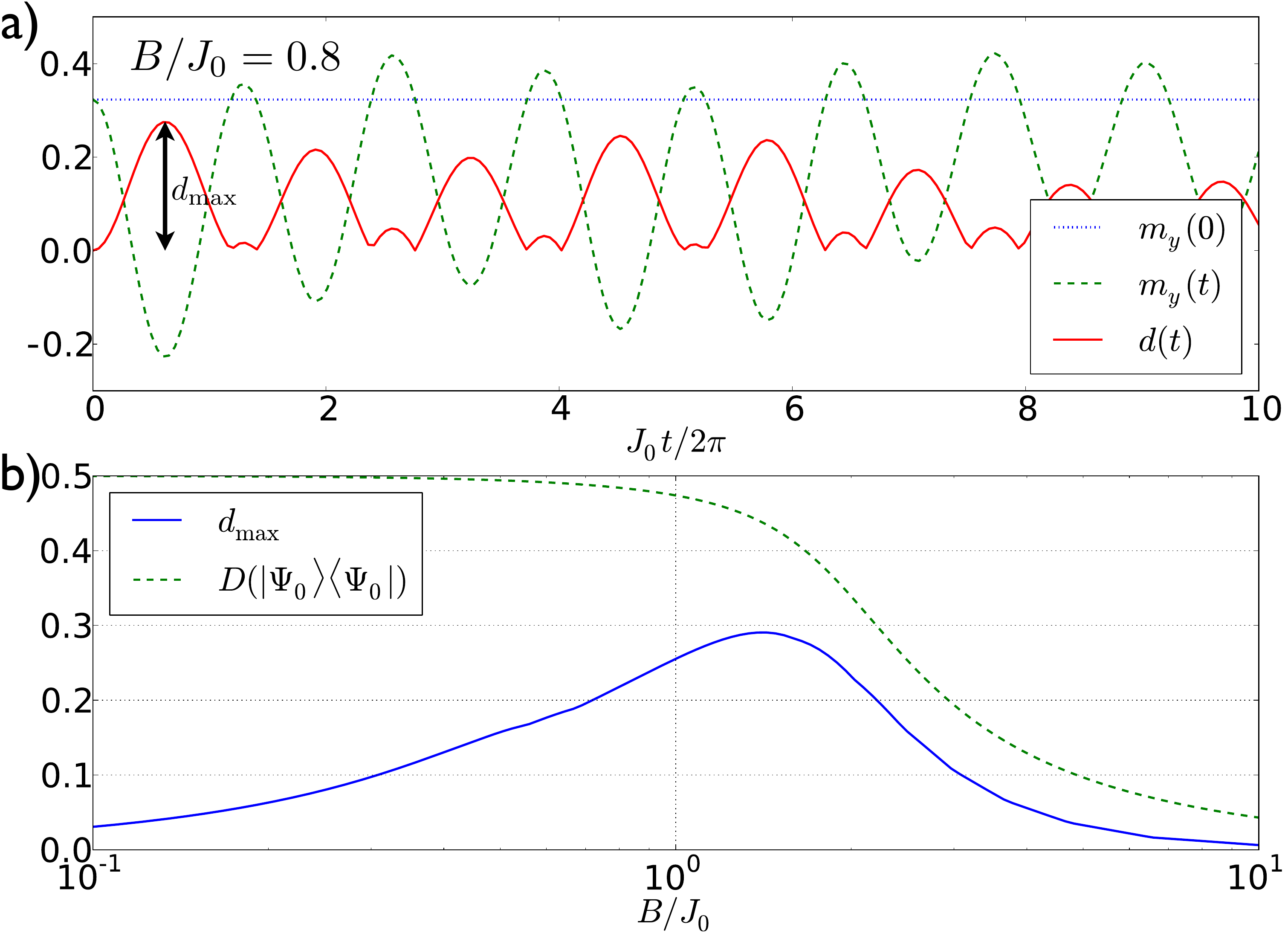}
\caption[Single-spin trace distance evolution shows a distinct peak at the quantum phase transition.]{Local dephasing of the ground state induces a dynamical evolution~(\ref{eq.dephindpolevo}) of the observed spin, thereby detecting and quantifying the ground-state entanglement between the measured spin and the rest of the chain, as quantified by the negativity~(\ref{eq.spinchainneg}). Further optimization of the local signal by maximization over all observed times $t$ discloses a finite-size precursor of the quantum phase transition in form of a pronounced peak in the vicinity of the critical point. Figure taken from \protect\cite{Gessner2014EPL}; M. Gessner \textit{et al.}, "Observing a quantum phase transition by measuring a single spin", Europhysics Letters, vol. \textbf{107}, issue 4, 2014, available at \url{http://iopscience.iop.org/article/10.1209/0295-5075/107/40005}.}
\label{fig.tracedistevoqpt}
\end{figure}

The local eigenbasis of each individual spin is for symmetry reasons ($Z_2$-symmetry: invariance of $H$ under a $\pi$-rotation around the $y$-axis) always given by the $y$-axis \cite{Gessner2014EPL,GessnerDiss}. The local dephasing is therefore always described by
\begin{align}\label{eq.localdephasingy}
\rho_{\Phi}=(\Phi\otimes\mathbb{I})|\Psi_0\rangle\langle\Psi_0|=\sum_{\varphi\in\{\uparrow_y,\downarrow_y\}}\left(|\varphi\rangle\langle \varphi|\otimes\mathbb{I}_B\right)|\Psi_0\rangle\langle\Psi_0|\left(|\varphi\rangle\langle\varphi|\otimes\mathbb{I}_B\right),
\end{align}
where $|\uparrow_y\rangle$ and $|\downarrow_y\rangle$ describe the eigenstates of $\sigma_y^{(1)}$ and, here, $\mathbb{I}_B$ is the identity operator on all remaining spins. Dephasing in this basis further yields the minimal trace distance~(\ref{eq.mindephdist}), as was shown in Section~\ref{sec.negativity}, and the dephasing disturbance thus coincides with the negativity \cite{Gessner2014EPL},
\begin{align}\label{eq.spinchainneg}
D(|\Psi_0\rangle\langle\Psi_0|)=\||\Psi_0\rangle\langle\Psi_0|-\rho_{\Phi}\|=\mathcal{N}(|\Psi_0\rangle\langle\Psi_0|).
\end{align}
The local evolution of the controllable spin is governed by
\begin{align}
\rho_A(t)=\mathrm{Tr}_B\{U(t)(\Phi\otimes\mathbb{I})|\Psi_0\rangle\langle\Psi_0|U^{\dagger}(t)\},
\end{align}
where $U(t)=e^{-iH_\alpha t/\hbar}$. At all times $t$, the local trace distance
\begin{align}\label{eq.tracedistqpt}
d(t)=\|\rho_A(t)-\rho_A(0)\|
\end{align}
yields a lower bound for the negativity $\mathcal{N}(|\Psi_0\rangle\langle\Psi_0|)$. This quantity is fully determined by the magnetization $m_y(t)=\mathrm{Tr}\{\rho_A(t)\sigma^{(1)}_y\}$ along the $y$ direction as \cite{Gessner2014EPL}
\begin{align}\label{eq.dephindpolevo}
d(t)=\frac{1}{2}|m_y(t)-m_y(0)|.
\end{align}
Again, we may take the time-maximum $d_{\max}$ of all local distances, Eq.~(\ref{eq.localmaxdist}), here coinciding with Eq.~(\ref{eq.minlocalwitness}), to obtain the strongest available lower bound on $\mathcal{N}$, as plotted in Fig.~\ref{fig.tracedistevoqpt}. 

A peak of the local signal around $B\simeq J_0$, indicates the quantum phase transition and hints at the position of the critical field. For larger values of $B$ both the total ground-state entanglement and the local signal decrease. For small values of $B$, entanglement is present, but not dynamically revealed. This can be understood through an analysis of the dephasing-induced excitations of the state $\rho_\Phi$ \cite{GessnerDiss}.

\subsubsection{Distribution of dephasing-induced excitations}
\begin{figure}[tb]
\centering
\includegraphics[width=.95\textwidth]{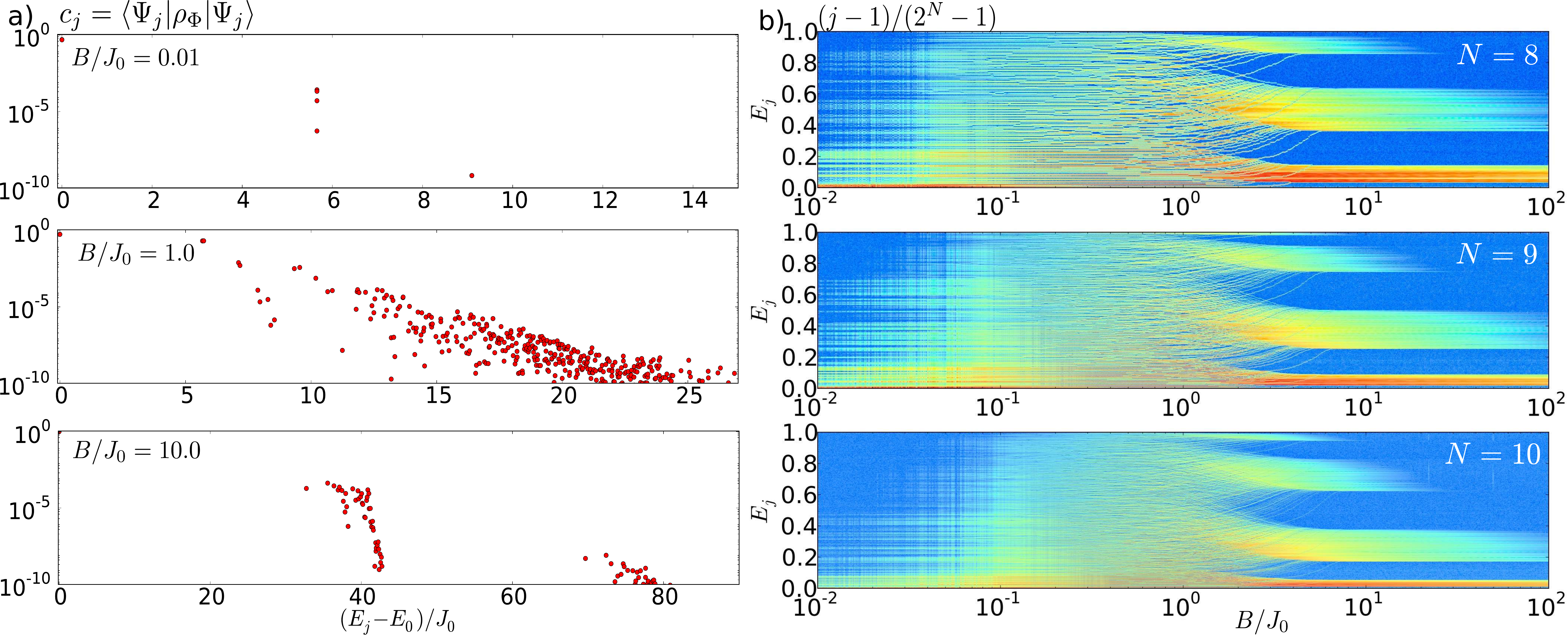}
\caption[Dephasing-induced excitations by local dephasing of the spin-chain ground state.]{Dephasing-induced excitations from the ground state for $\alpha=1$ and $N=10$ (left panels), as quantified by the overlap $c_j$ with energy eigenstates, Eq.~(\ref{eq.cj}). For small values of $B/J_0$ only few states above the ground state are populated, whereas around $B\simeq J_0$ a broad excitation spectrum is observed. For large values of $B/J_0$ regular bands, characteristic of the paramagnetic spectrum, are observed. The right panel shows the (renormalized) index of the excited states along $y$, where the color code is logarithmically scaled and normalized to 100 steps between the respective minimum and maximum values of $c_j$, increasing from blue to red. We observe quick convergence of the images with increasing $N$ towards an almost homogeneous excited-state distribution around the critical point of the quantum phase transition. Half of the paramagnetic bands are not populated for symmetry reasons (see text and Fig.~\ref{fig.spectrum5and8}). Figure adapted from \cite{GessnerDiss}.}
\label{fig.dephasingexcitations}
\end{figure}

The excitation spectrum of $\rho_{\Phi}$ is determined by the matrix elements of $\rho_\Phi$ in a basis of eigenstates $|\Psi_j\rangle$ of $H_\alpha$:
\begin{align}\label{eq.cj}
c_j=\langle\Psi_j|\rho_{\Phi}|\Psi_j\rangle.
\end{align}
Figure~\ref{fig.dephasingexcitations} shows the distribution of the dephasing-induced excitations $c_j$ on a logarithmic scale for different values of $B/J_0$, $J_0>0$. For small values of $B/J_0$ hardly any significant excited-state populations are created due to the local dephasing. In contrast, the intermediate regime $B/J_0\approx 1$ is characterized by a broadly distributed excitation spectrum. This shows, on the one hand, that the energy spectrum is widely spread, and on the other hand, that states of all energies are coupled to the ground state by means of the local dephasing. These features can be understood as the consequence of quantum chaotic structures \cite{Haake2001} that emerge close to the critical point in this model \cite{GessnerDiss}. For such dynamics, the local detection method is expected to be highly efficient since large parts of the state space are explored in the course of the dynamics, basically regardless of the initial condition \cite{Haake2001}, recall also Section~\ref{sec.efficacy}. For this reason, a complex dynamical evolution has a higher chance of successfully mapping the initial two-body coherences, responsible for discord, into the locally accessible subsystem. 

For large values of $B/J_0$ the energy depends linearly on $B$, with a slope proportional to the number of spins that orient along the $y$-direction, and thus the characteristic band structure of paramagnetic systems emerges around $E/J_0=-NB,-(N-2)B,-(N-4)B,\dots$. The finite width of these bands is due to nonzero values of $J_0$. However, this implies that besides those dephasing-induced populations at $(E-E_0)/J_0=40,80$ which are observed in Fig.~\ref{fig.dephasingexcitations} there are other bands around $(E-E_0)/J_0=20,60$ which are not reached by the local dephasing. The reason for this is again found in the $Z_2$-symmetry of the Hamiltonian: The ground state, being member of the even-parity subspace, cannot be mapped onto states of the odd-parity subspace by local dephasing~(\ref{eq.localdephasingy}), since this operation commutes with the parity operator \cite{GessnerDiss}. Hence, when applying local dephasing to the ground state, we remain in the parity subspace of the ground state. 

\begin{figure}[tb]
\centering
\includegraphics[width=.95\textwidth]{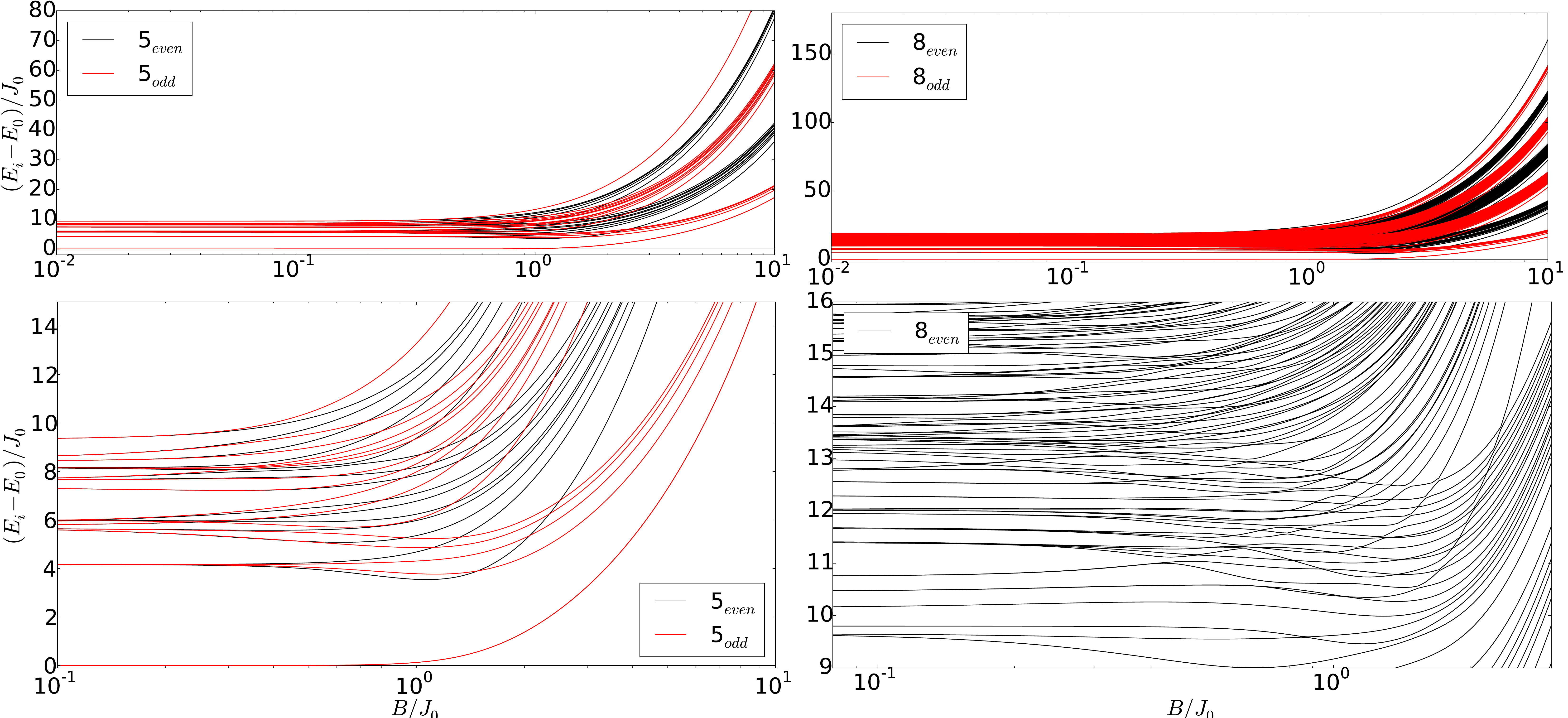}
\caption[Spectra of small spin systems separated into parity subspaces.]{Spectra of the spin chain Hamiltonian~(\ref{eq.spinchainepl}) for $\alpha=1$, $J_0>0$ and $N=5$ (left) and $N=8$ (right), respectively. Spectral lines corresponding to the two parity subspaces are distinguished by color, as indicated in the legends. As is seen from the top panels, the paramagnetic energy bands comprise only states of a definite parity subspace, causing the local dephasing operation to leave half of them unpopulated. We further find broadly distributed spectra around $B\simeq J_0$, even if restricting to only one of the two subspaces. Figure taken from \cite{GessnerDiss}.}
\label{fig.spectrum5and8}
\end{figure}

\begin{figure}[tb]
\centering
\includegraphics[width=.5\textwidth]{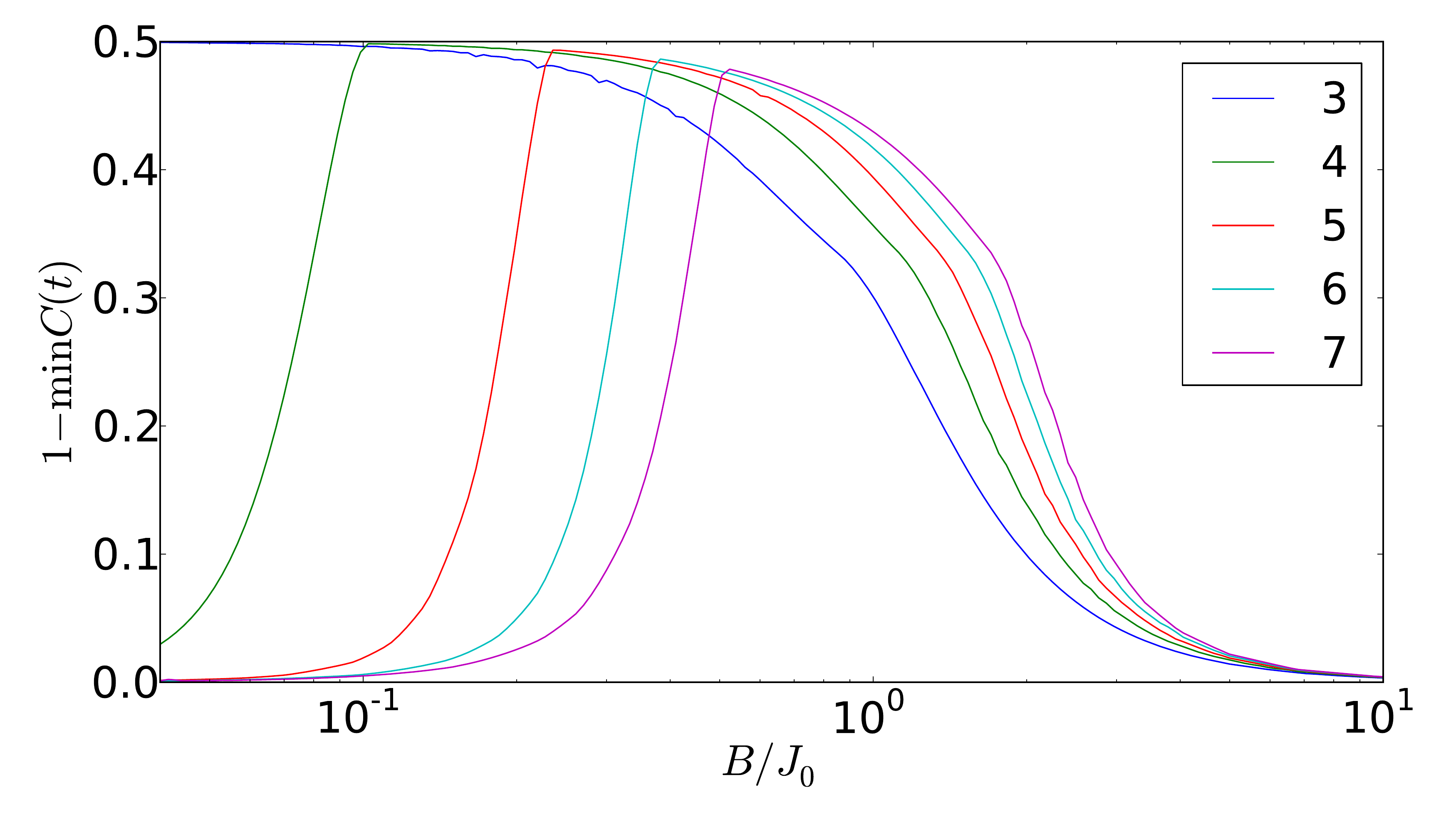}
\caption[Global time-autocorrelation function of the dephased state.]{Deviations of the dephased state $\rho'$ from the original state $\rho$ are quantified using the global time-autocorrelation function $C(t)$. Plotting the minimum value (over all $t$) as a function of $B/J_0$ confirms that no evolution takes place when $B$ is very large or very small. The plots display different values of $N$ (see legend) with $J_0>0$ and $\alpha=1$. Hence, the local signal, as shown in Fig.~\ref{fig.tracedistevoqpt} captures the qualitative behavior of the global dynamics, and little information is lost through the partial trace operation. Figure taken from \cite{GessnerDiss}.}
\label{fig.autocorr}
\end{figure}

Two ingredients are necessary for a successful mapping of the two-body coherences of $|\Psi_0\rangle$ to $\rho'_A(t_0)$: So far we have discussed the crucial aspect of populating a family of excited states such that $\rho'(t)=U(t)\rho_{\Phi}U^{\dagger}(t)$ experiences a suitable dynamics. However, this condition is not sufficient, since the partial trace operation may not disclose this dynamics to the observable subsystem. To see whether the state $\rho'(t)$ actually shows richer dynamics than $\rho'_A(t)$ in the case of $B\ll J_0$, we may consult the global time-autocorrelation function
\begin{align}
C(t)=\frac{1}{\mathcal{P}(\rho_{\Phi})}\mathrm{Tr}\{\rho_{\Phi}U(t)\rho_{\Phi}U^{\dagger}(t)\},
\end{align}
which is normalized by the purity $\mathcal{P}(\rho_{\Phi})=\mathrm{Tr}\rho_{\Phi}^2$ such that $C(0)=1$. The time evolution of $\rho'(t)$ depends on the coherences of $\rho_{\Phi}$ in the energy eigenbasis, as seen from the expression
\begin{align}
C(t)=\frac{1}{\mathcal{P}(\rho_{\Phi})}\sum_{ij}|\langle\Psi_j|\rho_{\Phi}|\Psi_i\rangle|^2e^{-i(E_i-E_j)t/\hbar}.
\end{align}
However, as Fig.~\ref{fig.autocorr} shows, the fact that no dynamical witness for entanglement is obtained around $B\ll J_0$ is not due to the local observation of the quantum system. In this parameter regime, hardly any dynamical evolution of $\rho'(t)$ can be observed \cite{GessnerDiss}.

\subsubsection{Thermal states: Local bound for the minimum entanglement potential}
\begin{figure}[tb]
\centering
\includegraphics[width=.6\textwidth]{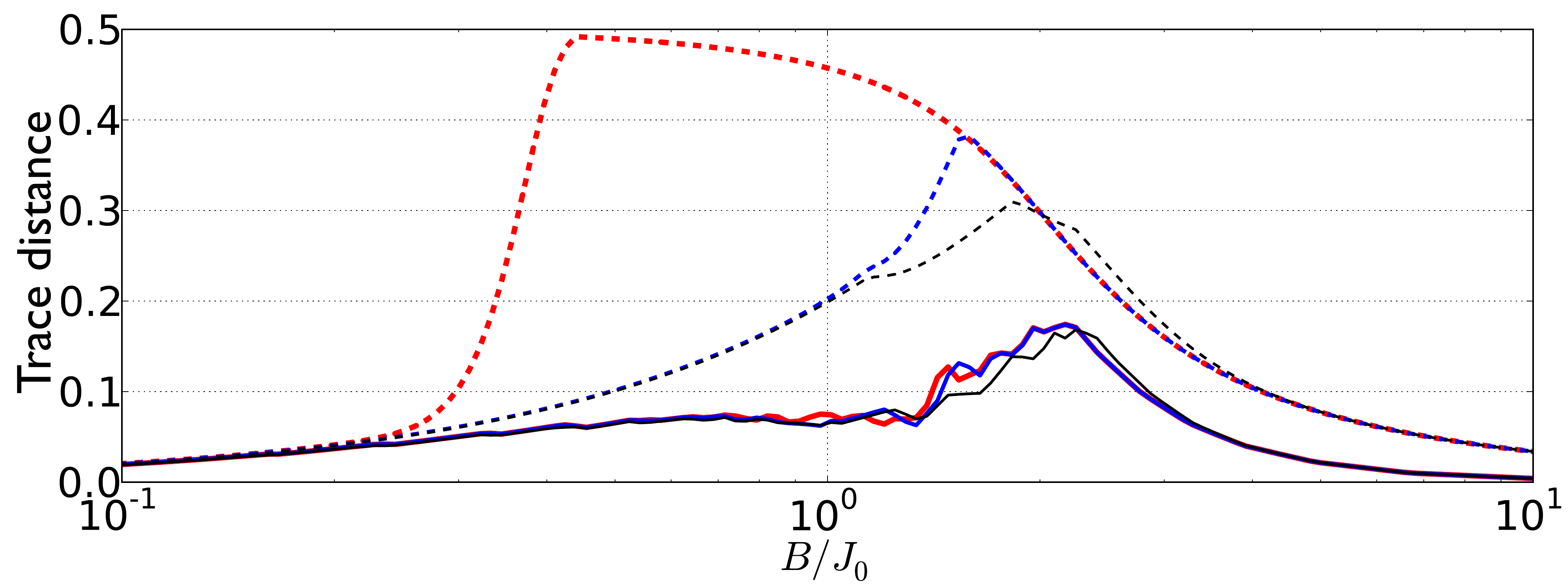}
\caption[Thermal discord-type correlations of low-temperature states and their single-spin signature.]{The minimum dephasing disturbance $D_{\min}$ (dashed lines), Eq.~(\ref{eq.mindephdist}), decreases dramatically when the thermal energy $kT$ exceeds the energy gap between the two states of lowest energy is smaller than the thermal energy $kT$. The local signal $d_{\min}$ (continuous lines), Eq.~(\ref{eq.minlocalwitness}), is more robust to increasing temperatures. Parameters are $N=7$ with $kT=10^{-5}$ (thick, red lines), $kT=0.1$ (medium, blue), and $kT=1$ (black, thin). Figure adapted from \protect\cite{Gessner2014EPL}; M. Gessner \textit{et al.}, "Observing a quantum phase transition by measuring a single spin", Europhysics Letters, vol. \textbf{107}, issue 4, 2014, available at \url{http://iopscience.iop.org/article/10.1209/0295-5075/107/40005}.}
\label{fig.tracedistqpttherm}
\end{figure}

Below the critical point, the energy gap between the ground state and the first excited state decreases rapidly with increasing $N$. Rather than a preparation of the pure ground state, one would, in realistic conditions, therefore expect to find a thermal state,
\begin{align}\label{eq.thermalstateqptepl}
\rho_{\beta}=\frac{e^{-\beta H}}{\mathrm{Tr}e^{-\beta H}},
\end{align}
with inverse temperature $\beta=1/kT$. When the energy gap, which also decreases with decreasing $B$ for fixed $N$, becomes smaller than the thermal energy $kT$, then the two neighboring states are mixed incoherently. As a consequence, all quantum correlations which are present in $|\Psi_0\rangle$ are removed in $\rho_{\beta}$, which can be shown to have zero discord \cite{GessnerDiss}. Thus, the minimal dephasing disturbance $D_{\min}$, as introduced in equation~(\ref{eq.mindephdist}), approaches zero when $B$ is decreased below a temperature-dependent value. Far away from any degeneracy it reduces, as expected, to the negativity of the energetically lower-lying state, see Fig.~\ref{fig.tracedistqpttherm}. 

To obtain a local bound on $D_{\min}$, one uses the optimized local witness, described in Eq.~(\ref{eq.minlocalwitness}). As seen in Fig.~\ref{fig.tracedistqpttherm}, this quantity is less sensitive to the mixing process than the total correlations, and a maximum signal can still be observed around the critical point.

%\begin{figure}[tb]
%\centering
%\includegraphics[width=.6\textwidth]{SpectrumAndGap}
%\caption[Lower part of the excitation spectrum and energy gap between ground states of the spin chain.]{Lower part of the excitation spectrum of the ferromagnetic spin chain~(\ref{eq.spinchainepl}) at $\alpha=1$ for different system sizes. The gap $\Delta$ between the ground- and first excited state (inset) determines at which temperature the low-temperature thermal state~(\ref{eq.thermalstateqptepl}) turns from a pure ground state to an incoherent mixture of both ground states. Unlike the pure ground state, which contains GHZ-type entanglement~(\ref{eq.ghzstate}), the incoherent mixture~(\ref{eq.thermalgroundstate}) contains neither entanglement nor discord-type correlations. In the thermodynamic limit the two ground states are degenerate for values of $B$ below the critical field. A detailed discussion of the spectrum is provided in chapter~\ref{sec.spinchainQPT}.}
%\label{fig.spectrumandgap}
%\end{figure}

\subsection{Atom-photon correlations during spontaneous emission}
Let us finally discuss an example of a dynamical system where the local detection method is unable to reveal initial entanglement in the subdynamics. In the spontaneous emission processes, atom and field start and end in factorized conditions, while the intermediate state contains atom-field entanglement. In a theoretical study reported in \cite{GessnerDiss} the local detection method was applied to such an entangled intermediate state using the atomic two-level system as the accessible subsystem, while the spontaneous emission process into the electromagnetic field modes provides the interacting dynamics.

The situation differs conceptually from the photonic experiment reported in Section~\ref{sec.photonicexperiment}, since the interaction here exchanges energy instead of only imprinting a relative phase. Moreover, due to the irreversibility of the process, the energy which is transmitted from the atom to the photons is irretrievably lost. In the trapped ion experiment, reported in Section~\ref{sec.localdetectionwithion}, energy was also exchanged between the two systems, but the ancilla system was described by a single mode instead of a continuum of modes, with the possibility of feedback from the environment to the controllable system.

The spontaneous emission process can be described by a unitary evolution $U(t)=e^{-iHt/\hbar}$, generated by the Hamiltonian $H=H_0+V$ with \cite{Weisskopf1930,Cohen-Tannoudji1992}
\begin{align}
H_0=E_e|e\rangle\langle e|+E_g|g\rangle\langle g|+\sum_{\mathbf{k}}\hbar\omega_{\mathbf{k}}a^{\dagger}_{\mathbf{k}}a_{\mathbf{k}}
\end{align}
and
\begin{align}
V=\sum_{\mathbf{k}}(g_{\mathbf{k}}a^{\dagger}_{\mathbf{k}}|g\rangle\langle e|+g^*_{\mathbf{k}}a_{\mathbf{k}}|e\rangle\langle g|).
\end{align}
The modes of the electromagnetic field are labeled by $\mathbf{k}$ and $|\mathbf{k}\rangle$ denotes a one-photon state created by the bosonic operators $a^{\dagger}_{\mathbf{k}}$. The atom-field coupling strength is determined by the constants $g_{\mathbf{k}}$.

The atom is initially prepared in the excited state, while the field starts out in the vaccum. The initial state $|e,0\rangle$ evolves as
\begin{align}\label{eq.transientstate}
|\Psi(t_0)\rangle=u_{00}(t_0)|e,0\rangle+\sum_{\mathbf{k}}u_{\mathbf{k}0}(t_0)|g,\mathbf{k}\rangle,
\end{align}
with the matrix elements $u_{00}(t)=\langle e,0|U(t)|e,0 \rangle$ and $u_{\mathbf{k}0}(t)=u^*_{0\mathbf{k}}(-t)=\langle g,\mathbf{k}|U(t)|e,0 \rangle$. The local eigenbasis is readily found to be $\{|e\rangle,|g\rangle\}$, and local dephasing transforms the above entangled state into the classically correlated reference state,
\begin{align}
\rho'(t_0)=(\Phi\otimes\mathbb{I})|\Psi(t_0)\rangle\langle\Psi(t_0)|.
\end{align}
The difference between the two states,
\begin{align}\label{eq.rhorhoprimespontan}
\rho(t_0)-\rho'(t_0)=\sum_{\mathbf{k}}\left(u_{00}(t_0)u^*_{\mathbf{k}0}(t_0)|e,0\rangle\langle g,\mathbf{k}|+u^*_{00}(t_0)u_{\mathbf{k}0}(t_0)|g,\mathbf{k}\rangle\langle e,0|\right),
\end{align}
quantifies the atom-field negativity at time $t_0$. To determine the negativity explicitly, the matrix elements of the unitary time evolution are evaluated with the resolvent method \cite{Cohen-Tannoudji1992} and the continuum limit is performed; for details see \cite{GessnerDiss}. One finds \cite{GessnerDiss}
\begin{align}\label{eq.negativityevolution}
\mathcal{N}(|\Psi(t_0)\rangle\langle\Psi(t_0)|)&=c\sqrt{e^{-\Gamma t_0}(1-e^{-\Gamma t_0})},
\end{align}
where $c$ is a constant, independent of $t_0$, and $\Gamma$ is the spontaneous emission rate. This confirms the presence of atom-field entanglement in the intermediate states of the emission process.

The ensuing difference in the evolutions of the atomic system from $t_0$ to $t_1$ is described by
\begin{align}
\rho_A(t_1,t_0)-\rho'_A(t_1,t_0)&=\text{Tr}_B\{U(t_1-t_0)(\rho(t_0)-\rho'(t_0))U^{\dagger}(t_1-t_0)\}\notag\\
&=2\mathrm{Re}\left[\sum_{\mathbf{k}}u_{00}(t_0)u^*_{\mathbf{k}0}(t_0)u_{00}(t_1-t_0)u^*_{0\mathbf{k}}(t_1-t_0)\right]\sigma_z.\label{eq.signal}
\end{align}
Using the same techniques as before to evaluate this quantity, one finds it to be zero for all values of $t_0$ and $t_1$ \cite{GessnerDiss}. This shows that the atomic evolution is insensitive to a replacement of all quantum correlations by classical correlations at any intermediate time $t_0$. Hence, the atom-field correlations that are created during the spontaneous emission process cannot be detected using the same dynamical evolution. This is expected to change when modifications of the uniform exponentially decaying evolution, e.g., through higher-order corrections \cite{Cohen-Tannoudji1992} or a structured environment \cite{PhysRevA.33.3610,Lambr2000,BreuerPetruccione2006,RevModPhys.88.021002} are introduced.

\section{Conclusions}
In conclusion, the quantum discord of an interacting bipartite system can be probed with manageable overhead by using the local detection method. To do this, control can be limited to only one of the two correlated subsystems while the second system might even be completely unknown and inaccessible. The protocol requires the realization of local state tomography of the accessible system and a local dephasing operation, which may be realized by a non-selective local measurement. Due to the destructive nature of the measurement process, the protocol requires multiple copies of the initial state, as is common practice in quantum mechanical experiments.

By limiting access to one of the two subsystems, only a small Hilbert space of much lower dimension than the full quantum system, needs to be controlled. This permits the detection of discord in high-dimensional and infinite-dimensional systems, where full tomographic methods and the measurement of witness operators can no longer be realized. The method is further applicable in an open-system scenario, where a controllable quantum system couples to an environment which is generally difficult to access \cite{BreuerPetruccione2006}.

The efficacy of the method depends strongly on the dynamical behavior of the interacting system. A non-vanishing local dynamical signature of the initial discord is expected to be found generically for systems with complex, e.g., chaotic dynamics, exploring large parts of the state space. The case studies summarized in this article also show that for regular quantum optical model systems, the local detection method can be implemented successfully. In the context of system-environment dynamics, the photonic experiment reported in \cite{Tang2014} demonstrates that non-Markovian effects \cite{RevModPhys.88.021002} are not needed to achieve this: The evolution of the controllable subsystem was described by irreversible pure dephasing but a strong signature of the initial discord was recorded.

Conversely, the theoretical case study of the spontaneous emission process showed that a local dynamical signature of the atom-field entanglement cannot be recorded \cite{GessnerDiss}. In this extreme case the dynamics is no longer of a purely dephasing nature, but instead, excitations are decaying irretrievably from the controllable system into an environment. Thus, the examples discussed in this article suggest that, for dissipative dynamics, the local detection method relies on the presence of structure in the environment, such that excitations are indeed being exchanged both ways between the subsystems, as was the case in the trapped-ion experiment \cite{Gessner2014NP}.

Further uses of the local detection method lie in the analysis of large interacting many-body systems by means of a small ``quantum probe'' \cite{PhysRevA.88.012108,Gessner2014EPL,Haikka2014,Gessner2014NJP,PhysRevA.92.010302,GessnerDiss}. This was illustrated in the context of a quantum phase transition, where a strong dynamical signal of ground-state entanglement and thermal discord was observed in the vicinity of the critical point, with the measurements being limited to a single spin \cite{Gessner2014EPL}.

\section*{Acknowlegments}
MG thanks the German Academic Scholarship Foundation (Studienstiftung des deutschen Volkes) for support during the work on his PhD thesis. HPB \& AB acknowledge support by the EU Collaborative project QuProCS (Grant Agreement 641277). AB thanks Dieter Jaksch and his group, as well as Keble College, for the hospitality he enjoyed during a research visit to Oxford.

\end{document}